\documentclass[aps,pra,twocolumn,superscriptaddress,amsmath,amssymb,floatfix]{revtex4-2}
\usepackage{graphicx}
\usepackage{amsthm}
\usepackage{booktabs}
\usepackage{xcolor}
\usepackage{tikz}
\usepackage{hyperref}
\hypersetup{pdftitle={Computational Complexity and Energy Landscapes of Coulomb Crystals},
  pdfauthor={Paco Navarro and Michael Emmerich}}

\newtheorem{theorem}{Theorem}
\newtheorem{lemma}{Lemma}
\newtheorem{proposition}{Proposition}
\newtheorem{corollary}{Corollary}
\theoremstyle{definition}
\newtheorem{definition}{Definition}
\newtheorem{remark}{Remark}

\newcommand{\R}{\mathbb{R}}
\newcommand{\Q}{\mathbb{Q}}
\newcommand{\Z}{\mathbb{Z}}
\newcommand{\eps}{\varepsilon}
\newcommand{\Coul}{\textsf{Coulomb-Subset}}

\begin{document}

\title{Computational Complexity and Energy Landscapes of Coulomb Crystals}

\author{Paco Navarro}
\affiliation{Department of Physics, University of California, Santa Barbara, California 93106, USA}
\author{Michael Emmerich}
\affiliation{Faculty of Information Technology, University of Jyv\"askyl\"a, Finland}

\date{November 22, 2026}

\begin{abstract}
Simulated annealing is widely used to search for ground states of trapped-ion Coulomb crystals and charges on a sphere (the Thomson problem). We show that complexity depends on what the instance specifies. For identical ions in a fixed trap, the ion number alone specifies the system. With unary ion number and binary energy threshold, the decision problem lies in P/poly and cannot be NP-hard unless the polynomial hierarchy collapses. With a logarithmic precision parameter, approximate configuration search also admits a solution function in FP/poly. Both results apply to the Thomson problem. These nonuniform bounds provide no uniform search algorithm or guarantee of practical efficiency. In simulations of up to 100 ions, the number of distinct minima found and the time to reach the best-known minimum grow approximately exponentially. For ions pinned to prescribed candidate sites, minimizing Coulomb energy over occupied sites is NP-hard in two and three dimensions; a finite-precision promise version is NP-complete. All ions are identical, with bare Coulomb repulsion; hardness resides in the site positions alone. We extend Emmerich's reduction from the Ising problem on planar cubic graphs, established for Riesz potentials with $s\geq 2$, to Coulomb repulsion ($s=1$). The resulting landscapes have large barriers between nearly degenerate minima. Barrier trees resolve this structure, and simulated annealing becomes trapped above the ground state in the cube example. At laboratory parameters, with individually tunable well depths, the encoding survives numerically on a half-filled square lattice. For Doppler-cooled strontium ions, this requires stiff wells a few to tens of micrometers apart, as in electrode trap arrays.
\end{abstract}

\maketitle

\section{Introduction}
\label{sec:intro}

Laser-cooled ions in radio-frequency and Penning traps crystallize into Coulomb crystals whose structure is set by the balance between the confining potential and the mutual Coulomb repulsion~\cite{DubinONeil99,Drewsen15,Thompson15}. The structure matters for quantum simulation and computation with large ion registers, for studies of structural phase transitions and friction~\cite{Kiethe17,Ruffert26}, and for sympathetic cooling and spectroscopy of molecular ions co-trapped with laser-cooled atomic ions, which now reaches single molecules with rotational resolution~\cite{calvin2023single,PhysRevLett.134.153002}.

Finding the ground-state structure is routinely treated as a hard global-optimization problem. Molecular-dynamics simulations combined with simulated annealing are the standard tool~\cite{Kirkpatrick83,Ludwig05}: Kiesenhofer \emph{et al.} anneal each parameter set 500 times to collect the low-lying configurations of two-dimensional crystals of more than 100 ions~\cite{Kiesenhofer23}, and R\"uffert and Mehlst\"aubler define the ground state of a three-dimensional crystal operationally, as the lowest configuration recovered at least ten times~\cite{Ruffert26}. The same difficulty appears in the Thomson problem of $N$ charges on a sphere~\cite{Thomson1904,Smale98}, which has been attacked with simulated annealing since 1985~\cite{WV85b} and whose number of local minima is observed numerically to grow exponentially with $N$~\cite{ErberHockney91,WalesUlker06,Calef15,Amore25,Stillinger99}.

This difficulty is often attributed to computational complexity. For instance, the Thomson problem has been described as ``known to be NP-hard''~\cite{AmoreZarate23}, with a citation to Wille and Vennik~\cite{WV85,WV85b}. Wille and Vennik, however, proved hardness for clusters on a lattice with site-dependent weights, and Adib later showed that their argument does not apply to pair potentials that depend only on distance~\cite{Adib05}; Adib's own hardness result uses a step potential supplied with the input. Closer to the present setting, Adamson \emph{et al.} proved NP-hardness of removing ions from a given finite arrangement so as to minimize its energy, for Buckingham--Coulomb potentials with short-range parameters supplied with the input and, for the Coulomb potential alone, with charges of both signs and unbounded magnitude, by reduction from knapsack~\cite{ADGP21}; a periodic version on $\Z^d$ is NP-hard for a class of finite-range pair potentials whose parameters are part of the input~\cite{ADGP22}. In these results the hardness is carried by the charges or the force-field parameters, not by the positions alone.
Recent work showed that selecting $k$ points of minimum Riesz energy $\sum r^{-s}$ is NP-hard in general metric spaces, in the plane when $s$ is part of the input~\cite{EPD26}, and, by Emmerich, in the plane for the fixed exponent $s=2$~\cite{Emm26}. Ref.~\cite{Emm26} also notes that the argument extends to fixed $s>2$, where the tails decay faster. On the line, the problem is solvable in polynomial time~\cite{EmmLine26}. For identical charges, the Coulomb exponent $s=1$, the case relevant to ions, was not covered: Ref.~\cite{EPD26} lists fixed exponents as an open problem, and below $s=2$ the tail estimates of Ref.~\cite{Emm26} become harder rather than easier.

Here we ask what these results imply for ion crystals, separating two kinds of problem (Fig.~\ref{fig:freepinned}).

\begin{itemize}
\begin{figure*}[t]
\includegraphics[width=\textwidth]{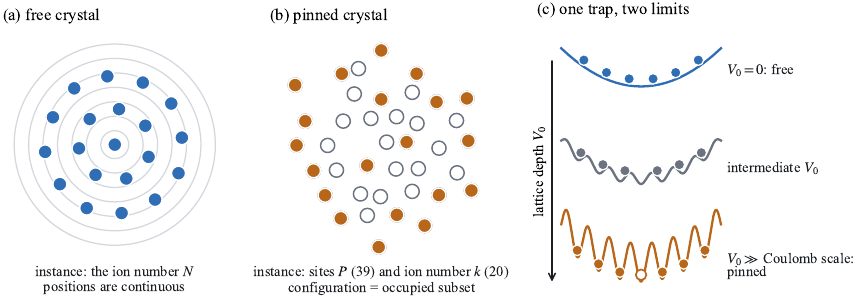}
\caption{Free and pinned ion crystals (Definition~\ref{def:crystals}). (a) A free crystal: the lowest-energy configuration found for $N=20$ ions in a two-dimensional harmonic trap (circles are equipotentials). The ion positions are continuous, and the problem instance is the number $N$ alone. (b) A pinned crystal: the potential provides a prescribed set $P$ of wells (open circles), and a configuration is the choice of which $k$ wells are occupied (filled; shown is the lowest-energy choice found for $k=20$ ions on 39 sites). The instance now carries the whole site set. (c) Both are limits of one physical situation, a smooth trap plus a lattice potential of depth $V_0$. Without the lattice the ions sit where the Coulomb repulsion places them; for a lattice much deeper than the Coulomb energy scale they are locked into wells, and the only freedom left is which wells to fill (one well is left empty here).}
\label{fig:freepinned}
\end{figure*}
\item \emph{Free crystals.} For identical ions in a fixed trap, the physical system is specified by the ion number. With unary ion count and a binary energy threshold, an elementary advice argument gives membership in P/poly, so it is not NP-hard unless the polynomial hierarchy collapses (Theorem~\ref{thm:free}); the same argument covers the decision version of the Thomson problem. With unary ion count and precision parameter, Approximate Coulomb Configuration Search (ACCS) on the fixed sphere or in a fixed positive-definite harmonic trap also admits a solution function in $\mathrm{FP/poly}$ (Proposition~\ref{prop:accs}). These elementary bounds clarify the encoding without providing uniform search algorithms; we quantify practical difficulty separately (Sec.~\ref{sec:practice}).

\item \emph{Pinned crystals.} When the candidate positions are part of the input, the problem is NP-hard in two and three dimensions (Theorem~\ref{thm:main}). Its finite-precision version is NP-complete (Theorem~\ref{thm:npc}); whether the exact problem lies in NP is open, because checking a configuration exactly means comparing sums of square roots. The source encoding, selectors, wires, objective boxes, and field-compensation mechanism come from Emmerich's $s=2$ reduction~\cite{Emm26}. To establish $s=1$, we derive the Coulomb angular law and quartic field term, strengthen the estimates for coupling perturbations and tree consistency, and choose a rational threshold for the irrational energy scale (Sec.~\ref{sec:s1}). We explain the construction in physical terms in Sec.~\ref{sec:pinned} and collect the $s=1$ estimates in Appendix~\ref{app:proof}.

\item \emph{Landscape.} For disjoint copies of $K_4$, the number of nearly degenerate minima grows exponentially with the number of source vertices, and large barriers separate the minima (Corollary~\ref{cor:landscape}). We further study these barriers using barrier trees and simulated annealing on the cube graph (Sec.~\ref{sec:landscape}).
\item \emph{Laboratory parameters.} The proof's gadgets are far smaller than any realizable well spacing. Numerically, the encoding nevertheless survives when the sites form a half-filled square lattice, provided the well depths are individually tunable; for Doppler-cooled Sr$^+$ this requires stiff wells a few to tens of micrometers apart, as in electrode trap arrays.
\end{itemize}

The logical skeleton of the reduction is machine-checked in Lean~4 (Appendix~\ref{app:lean}); the geometric estimates are proved by hand, and the tightest of them are certified by interval arithmetic.

\section{Free and pinned crystals}
\label{sec:free}

The complexity of a ground-state problem depends on what the problem instance specifies. We distinguish two idealizations of an ion crystal (Fig.~\ref{fig:freepinned}).

\begin{definition}[Free and pinned crystals]
\label{def:crystals}
Consider $N$ ions in $\R^d$, held by an external potential $U$ and repelling one another through the Coulomb interaction.

(i) The crystal is \emph{free} if $U$ is a fixed, smooth confining potential that is the same for every instance, such as the pseudopotential of a given Paul trap. The ion positions are continuous variables, and an instance is specified by the ion number alone (and, for a mixed crystal, by the numbers of ions of each of a fixed set of species).

(ii) The crystal is \emph{pinned} if $U$ consists of a prescribed finite set $P\subset\R^d$ of wells that are deep and narrow compared with all Coulomb energies and spacings, and $P$ is part of the instance. Each ion then sits at the bottom of a well, no well holds more than one ion, and a configuration is the set $S\subseteq P$ of occupied wells.
\end{definition}

The two notions are opposite limits of one physical situation [Fig.~\ref{fig:freepinned}(c)]. In a free crystal the structural information comes from the interaction itself, whereas in a pinned crystal it is supplied by the landscape. This difference determines the complexity.

\subsection{Elementary nonuniform bounds}

We record elementary nonuniform bounds for fixed free-crystal families. Their purpose is to delimit NP-hardness claims under the stated input encodings; they do not provide algorithms for finding ground states.

Fix the dimension $d$ and a continuous confining potential $U:\R^d\to\R$, with $U(x)\to\infty$ as $|x|\to\infty$. For $N$ identical ions, let
\begin{equation}
E_N(x_1,\dots,x_N)=\sum_{i=1}^N U(x_i)+\sum_{i<j}\frac{1}{|x_i-x_j|},
\label{eq:EN}
\end{equation}
in units where $q^2/4\pi\eps_0=1$, and write $E^\star(N)=\inf E_N$. The decision language is
\[
L_U=\{\langle1^N,\tau\rangle:\tau\in\Q,\ E^\star(N)\le\tau\},
\]
with $N$ in unary and $\tau$ in binary. Thus input length is measured in $N$ and the bit length of $\tau$; the trap is not part of the input.

A decision problem is in $\mathrm{P/poly}$ if a polynomial-time algorithm can solve it using a polynomial-length \emph{advice string}, shared by all inputs of the same length~\cite{AroraBarak}. The advice may depend on that length but not on the particular input, and no algorithm for constructing it is required.

\begin{theorem}[Elementary decision bound]
\label{thm:free}
For every fixed dimension $d$ and fixed confining potential $U$, $L_U\in\mathrm{P/poly}$. If $L_U$ is NP-hard under polynomial-time Turing reductions, then $\mathrm{NP}\subseteq\mathrm{P/poly}$ and $\mathrm{PH}=\Sigma_2^p$.
\end{theorem}
\begin{proof}
Let $\Q_n$ be the rationals with binary encodings of at most $n$ bits. For each $N\le n$, store the least $a_N\in\Q_n$ satisfying $E^\star(N)\le a_N$, or $\bot$ if none exists. This table has $O(n^2)$ bits. For a valid input $\langle1^N,\tau\rangle$ of length $n$, one has $\tau\in\Q_n$ and
\[
E^\star(N)\le\tau\quad\Longleftrightarrow\quad
a_N\ne\bot\ \text{and}\ a_N\le\tau.
\]
A lookup and a rational comparison therefore suffice; malformed inputs are rejected. If SAT reduces to $L_U$ in polynomial time, concatenate the advice tables for all polynomially bounded query lengths to obtain $\mathrm{SAT}\in\mathrm{P/poly}$. The Karp--Lipton theorem then gives $\mathrm{PH}=\Sigma_2^p$~\cite{KL80}.
\end{proof}

The table already encodes the threshold answers; the proof gives no procedure for computing its entries. In fact, the same argument works with \emph{any} fixed sequence $f(N)$ in place of $E^\star(N)$, whether computable or not. It therefore also covers the Thomson decision problem on the fixed unit sphere, and extends to a fixed number of ion species with unary counts. Assuming that the polynomial hierarchy does not collapse, these decision problems are not NP-hard under the stated reductions. This qualifies NP-hardness claims such as Ref.~\cite{AmoreZarate23} for this encoding.

\begin{remark}\label{rem:encoding}
Unary $N$ is a modeling choice, compatible with explicitly listing $dN$ coordinates. With binary $N$, the advice table can be exponentially large. An explicit configuration can also have exponentially many coordinates and does not automatically give a polynomial-size NP certificate. The argument does not classify this encoding or models in which the trap varies with the input.
\end{remark}

\emph{Approximate configuration search.} In Approximate Coulomb Configuration Search (ACCS), the input is $(1^N,1^q)$ and the output is a rational configuration of energy at most $E^\star(N)+2^{-q}$. Both $N$ and the precision parameter $q$ are unary, so input length is $\Theta(N+q)$. On the fixed sphere, output points must lie exactly on $S^2$. The additional issue is whether sufficiently accurate configurations have polynomial-length descriptions.

\begin{proposition}[Elementary search bound]
\label{prop:accs}
For the Thomson problem on the fixed unit sphere $S^2$, or a fixed harmonic trap $U(x)=\tfrac12x^\top A x$ in fixed dimension $d$, with fixed symmetric positive-definite $A\in\Q^{d\times d}$, ACCS admits a solution function in $\mathrm{FP/poly}$: polynomial time with polynomial-length advice.
\end{proposition}
\begin{proof}
We first bound the description length of a configuration with energy error at most $2^{-m}$. For $N\ge2$, set $B=\binom N2$.

\emph{Sphere.} Two independent uniform points on $S^2$ have mean Coulomb energy $1$, so $E^\star(N)\le B$. Every pair at a minimizer is therefore at least $B^{-1}$ apart. Moving each point by at most $\eta\le(4B)^{-1}$ changes the total pair energy by at most $4B^3\eta$. Rational points on the sphere approximate any point to distance $\eta$ using $O(\log(1/\eta))$ bits: round the bounded parameters $|u|,|v|\le1$ in the two stereographic maps
\[
(u,v)\longmapsto\frac{(2u,\,2v,\,\pm(1-u^2-v^2))}{1+u^2+v^2}.
\]
These maps cover the sphere with uniformly bounded derivatives on the parameter squares. Choosing $\eta=2^{-m}/(4B^3)$ gives the required error with $O(N(m+\log N))$ bits.

\emph{Harmonic trap.} Let $\lambda,\Lambda>0$ be the extreme eigenvalues of $A$, and put $C_N=(\Lambda+1)N^3$. The trial points $x_i=i e_1$ give $E^\star(N)\le C_N$. Positivity of the energy terms then gives, at a minimizer,
\[
|x_i|\le R_N:=\sqrt{2C_N/\lambda},\qquad
|x_i-x_j|\ge C_N^{-1}.
\]
For displacements of norm at most $\eta\le(4C_N)^{-1}$, the pair-energy change is at most $4BC_N^2\eta$, and the trap-energy change at most $N\Lambda(R_N\eta+\eta^2/2)$. Thus the total change is at most $L_N\eta$, where
\[
L_N=4BC_N^2+N\Lambda(R_N+1)=O(N^8).
\]
Since $L_N\ge4C_N$, dyadic rounding within $\eta=2^{-m}/L_N$ is admissible and gives error at most $2^{-m}$. The radius and precision bounds again require $O(N(m+\log N))$ bits, with constants depending on the fixed trap and dimension. For $N=1$, use a fixed rational point on the sphere or the trap origin.

For input length $m$, the advice stores one such configuration for every $N\le m$, using $O(m^3)$ bits in total. On input $(1^N,1^q)$, return the stored configuration for $N$. Its error is at most $2^{-m}\le2^{-q}$.
\end{proof}

Here the advice explicitly contains approximate solutions. The approximation estimates justify their polynomial description length; the subsequent lookup is elementary. NP-hardness under polynomial-time oracle reductions that succeed for \emph{every} valid ACCS output would imply $\mathrm{NP}\subseteq\mathrm{P/poly}$ and hence $\mathrm{PH}=\Sigma_2^p$~\cite{KL80}. A uniform algorithm polynomial in $N+\log(1/\eps)$ remains unresolved by this argument, as does the search bound for general confining $U$. Smale's seventh problem, concerning uniform construction of near-minimal logarithmic-energy configurations, is likewise unaffected~\cite{Smale98}.

\emph{Comparison with quadratic programming.} Quadratic programming can be NP-hard with only one negative eigenvalue, but the constraints in that result vary with the instance~\cite{PardalosVavasis91}. For the Coulomb potential,
\[
\nabla^2|z|^{-1}=|z|^{-3}(3uu^\top-I_d),\qquad u=z/|z|.
\]
The transverse curvature is negative for $d\ge2$, but nonconvexity alone does not establish NP-hardness. In one dimension, each $(x_j-x_i)^{-1}$ is convex on $x_1<\cdots<x_N$, so convex $U$ gives a convex energy there. The pinned problem supplies varying instance data through the prescribed site positions (Sec.~\ref{sec:pinned}).

Our results neither establish nor rule out NP-hardness when a rational positive-definite trap matrix $A$ and spatial dimension $d$ are part of the input.

\subsection{How hard are free crystals to compute in practice?}
\label{sec:practice}

Theorem~\ref{thm:free} conditionally excludes NP-hardness for the stated decision problem, but does not make finding free-crystal ground states easy. A natural measure is the \emph{time to solution} (TTS) of a randomized search~\cite{Ronnow14}. If one attempt (a local minimization from a random start, or one annealing run) costs time $t_1(N)$ and reaches the ground state with probability $p(N)$, the expected time to find it is
\begin{equation}
\mathrm{TTS}(N)=\frac{t_1(N)}{p(N)}.
\label{eq:tts}
\end{equation}
We estimate both quantities from repeated runs. Since the true ground state is not known, we measure $p(N)$ against the \emph{best-known minimum}, the lowest energy found for each $N$ over all $4000$ local minimizations and $160$ annealing runs; a run counts as successful if its energy agrees with this minimum to relative tolerance $10^{-7}$.

\begin{figure}[t]
\includegraphics[width=\columnwidth]{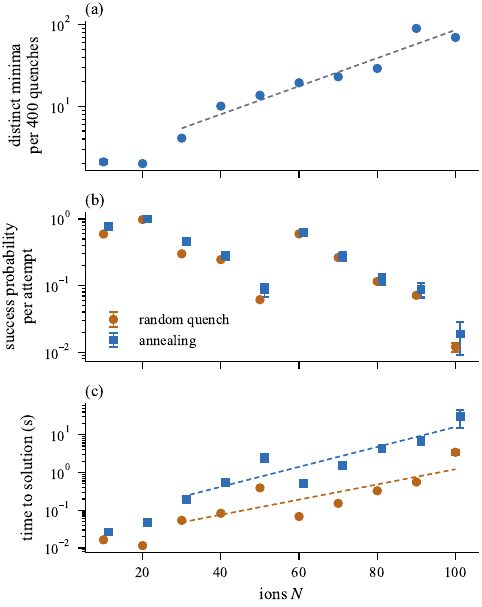}
\caption{Practical difficulty of free crystals: $N$ ions in a two-dimensional isotropic harmonic trap, $E=\sum_i r_i^2/2+\sum_{i<j}r_{ij}^{-1}$. The data comprise 10 independent batches per $N$, each with 400 local minimizations (L-BFGS) from random starts and 16 simulated-annealing runs. (a) Mean number of distinct minima found per batch. (b) Mean probability of reaching the best-known minimum, the lowest energy found over all attempts at that $N$, within relative tolerance $10^{-7}$; this is not a certified ground state. (c) Estimated time to solution, Eq.~\eqref{eq:tts}, computed as the mean time per attempt divided by the mean success probability. In (b,c), orange circles denote random quenches and blue squares denote annealing, slightly offset horizontally for visibility. Error bars in (a,b) are standard errors across batches; those in (c) propagate the uncertainty in the success probability, holding the mean attempt time fixed. Dashed curves are exponential fits for $N\ge30$.}
\label{fig:free}
\end{figure}

Figure~\ref{fig:free}(a,b) shows the sampled minima and success probabilities for $N\le100$ ions in a two-dimensional isotropic trap~\cite{BedanovPeeters94}; panel~(c) shows the corresponding estimates of time to solution. Over this range, the number of distinct minima found in a batch of 400 quenches, a lower bound on the number of metastable crystals, grows approximately as $e^{N/25}$ [Fig.~\ref{fig:free}(a)], and the probability that a local minimization from a random start reaches the best-known minimum falls from near unity to about $1\%$ at $N=100$ [Fig.~\ref{fig:free}(b)]. It does so non-monotonically, with ``magic'' ion numbers such as $N=60$ whose best-known minima have unusually large basins. The measured time per attempt grows roughly as $N^{0.7}$, so the decrease in $p$ accounts for most of the growth in time to solution. Over the investigated range it is well described by
\begin{equation}
\mathrm{TTS}(N)\approx\mathrm{poly}(N)\,e^{\kappa N},\qquad \kappa\approx0.046,
\label{eq:ttsfit}
\end{equation}
for random quenches, a tenfold increase every $\sim50$ ions; the annealing data give $\kappa\approx0.061$ [Fig.~\ref{fig:free}(c)]. Magic numbers scatter individual points by up to a factor of $\sim6$ about the fit. This is evidence for exponential scaling over $N\le100$, not a proof of asymptotic exponential scaling.

These observations explain why global stochastic searches are used. A single minimization succeeds with a probability that falls rapidly with $N$, consistent with a landscape of many nearly degenerate metastable crystals whose energy gaps shrink with $N$~\cite{Ludwig05,Amore25}. Simulated annealing~\cite{Kirkpatrick83} lets thermal fluctuations carry the system over the barriers between basins while the temperature is lowered, so that the search settles preferentially into the lowest basins. In our runs it reaches the best-known minimum more often per attempt than a random quench [Fig.~\ref{fig:free}(b)], but each attempt costs more, and both approaches scale approximately exponentially for $N\le100$. 

\section{Pinned crystals}
\label{sec:pinned}

\subsection{Problem and results}

For a pinned crystal [Definition~\ref{def:crystals}(ii)] with wells of equal depth, the energy of an occupation pattern $S\subseteq P$ is, up to a constant, the Coulomb energy of the occupied sites (unequal depths, or a smooth confining envelope as in Fig.~\ref{fig:freepinned}(c), would add a one-body term $\sum_{p\in S}U(p)$),
\begin{equation}
E_1(S)=\sum_{\{p,q\}\subseteq S}\frac{1}{|p-q|}.
\label{eq:E1}
\end{equation}
The subscript $1$ refers to the exponent $s=1$ of the Riesz energy $\sum|p-q|^{-s}$.

\begin{definition}[$\Coul_d$]
\label{def:coul}
Input: a finite set $P\subset\Q^d$ of candidate sites, an integer $k\le|P|$, and $\tau\in\Q$. Question: is there an occupation pattern $S\subseteq P$ with $|S|=k$ ions and $E_1(S)\le\tau$?
\end{definition}
\begin{figure*}[t]
\includegraphics[width=\textwidth]{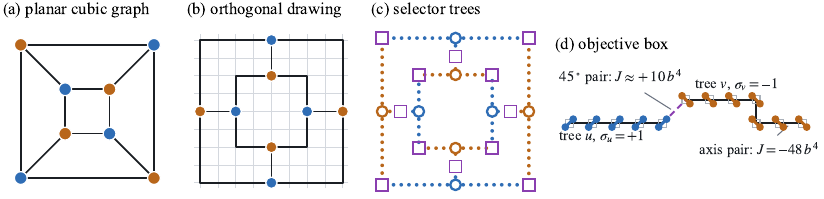}
\caption{The construction of Ref.~\cite{Emm26}, illustrated on the cube graph $Q_3$. (a) A planar cubic graph, colored by a ground state of the source energy $H_B$: blue vertices are selected ($\sigma_v=+1$) and form a maximum independent set; orange vertices are not selected. (b) A crossing-free orthogonal drawing on the integer grid. (c) Each vertex becomes a three-branch tree of selector gadgets at unit spacing (dots), and each edge is cut by an objective box (squares). (d) Detail of an objective box. Each selector is a square of four candidate sites with two ions on one diagonal; the diagonal encodes the bit. Axis-aligned neighbors couple ferromagnetically (solid lines) and copy the bit along a wire; the distinguished $45^\circ$ pair couples antiferromagnetically (dashed), producing the edge term $\lambda\sigma_u\sigma_v$. The displayed coupling values are leading-order approximations. Selector sizes are exaggerated.}
\label{fig:reduction}
\end{figure*}

\begin{theorem}[Pinned crystals are NP-hard]
\label{thm:main}
$\Coul_2$ is NP-hard under polynomial-time many-one reductions. Consequently $\Coul_d$ is NP-hard for every fixed $d\ge2$, in particular in three dimensions.
\end{theorem}

The theorem is the case $s=1$ of the construction of Ref.~\cite{Emm26}, which proves the analogous statement for $s=2$; we explain the construction in Secs.~\ref{sec:construction}--\ref{sec:farfield} and give the $s=1$ proof in Appendix~\ref{app:proof}. The extension to $d\ge 2$ follows by appending zero coordinates. On the line the problem is solvable in polynomial time~\cite{EmmLine26}, so unless $\mathrm{P}=\mathrm{NP}$ the hardness transition occurs between $d=1$ and $d=2$.

Is $\Coul_d$ also in NP? A ``yes'' answer has a short certificate, the set $S$, but checking it means comparing a sum of reciprocal square roots of rationals with $\tau$. This is an instance of the sum-of-square-roots problem, which is not known to be in NP~\cite{Garey76}; the best known upper bound gives $\Coul_d\in\mathrm{NP}^{\mathrm{PosSLP}}\subseteq\mathrm{CH}\subseteq\mathrm{PSPACE}$~\cite{Allender09}. At $s=2$ all energies are rational and this obstacle is absent~\cite{Emm26}. Physically, however, energies are only ever known to finite precision, and the finite-precision version is NP-complete. We use the standard conventions for promise problems~\cite{Goldreich06}, with all numbers encoded in binary. Because $E_1$ is homogeneous of degree $-1$, only the relative precision $\eps/\tau$ matters; in our instances it is inverse-polynomial in the input size. We make no claim about constant-factor approximation.

\begin{theorem}[Finite precision: NP-complete]
\label{thm:npc}
Consider the promise problem $\textsf{Gap-}\Coul_d$: the input is $(P,k,\tau,\eps)$ with $\eps>0$ rational; answer yes if some $|S|=k$ has $E_1(S)\le\tau$, and no if every $|S|=k$ has $E_1(S)>\tau+\eps$. For every $d\ge2$ this problem is NP-complete.
\end{theorem}

\begin{proof}
Membership: the certificate is $S$. For each of the $\binom{k}{2}<k^2/2$ pairs, compute a rational approximation of $1/|p-q|$ with absolute error at most $\eps/(2k^2)$ by Newton iteration, in time polynomial in the bit length of $|p-q|^2$ and in $\log(k^2/\eps)$. The approximate sum $\tilde E$ satisfies $|\tilde E-E_1(S)|<\eps/4$. Accept if $\tilde E\le\tau+\eps/2$: on a yes-instance with witness $S$, $\tilde E<\tau+\eps/4$; on a no-instance every $S$ has $\tilde E>\tau+3\eps/4$. Hardness: in the instances produced by the reduction, no-instances satisfy $E_1(S)>\tau+1.6\lambda$ for every $S$, with $\lambda>7b^4$ for an explicit rational $b$ (Appendix~\ref{sec:proof}). Setting $\eps=7b^4$ makes every produced instance satisfy the promise.
\end{proof}

\subsection{The construction of Ref.~\cite{Emm26}}
\label{sec:construction}

NP-hardness is proved by a \emph{reduction}: a polynomial-time procedure that converts any instance of a problem already known to be NP-hard into a pinned-crystal instance with the same yes/no answer. Following Ref.~\cite{Emm26}, the source problem is maximum independent set on planar cubic graphs~\cite{GJS76,Mohar01}: given a graph $G$ and an integer $K$, is there a set of $K$ vertices no two of which are joined by an edge?

The analogy is direct: chosen vertices may not be neighbors, and occupied sites repel. The obvious encoding places one site per vertex, puts adjacent vertices close together, and fills $K$ sites. It fails because the Coulomb interaction is long range. The $\binom{K}{2}$ distant pairs add an energy that depends on the whole configuration and can outweigh a single close pair, and making close pairs dominate would require nonadjacent vertices to be much farther apart than adjacent ones, which only disjoint unions of cliques allow.

Emmerich's idea is to encode each binary variable not in \emph{whether} a site is occupied but in \emph{how} a fixed number of ions is arranged~\cite{Emm26}. If both alternatives carry the same charge and dipole moment, they differ only at quadrupole order, and their influence on distant variables falls off much faster than the Coulomb interaction itself. The construction uses three gadgets, placed along a crossing-free drawing of the graph (Fig.~\ref{fig:reduction}):
\begin{itemize}
\item A \emph{selector} is a tiny square of four sites holding exactly two ions on one of its two diagonals. The choice of diagonal is a spin $\sigma=\pm1$.
\item A \emph{wire} is a chain of selectors at unit spacing. Neighboring selectors along an axis prefer the same diagonal, so a wire copies one spin over a long distance. Each vertex of the graph becomes a tree of three wires.
\item An \emph{objective box} joins the wires of two adjacent vertices at $45^\circ$, where neighboring selectors prefer opposite diagonals. This penalizes selecting both endpoints of an edge.
\end{itemize}
Written as spins, $\sigma_v=+1$ for a selected vertex, independent set becomes the ground-state problem of Barahona's Ising Hamiltonian~\cite{Bar82},
\begin{equation}
H_B(\sigma)=\sum_{uv\in E}\sigma_u\sigma_v+\sum_{v\in V}\sigma_v ,
\label{eq:HB}
\end{equation}
whose minimum lies at or below a threshold $H_0$ if $G$ has an independent set of size $K$, and at least $H_0+4$ otherwise (Appendix~\ref{app:source}). The pinned crystal must reproduce $\lambda H_B$ for some energy unit $\lambda>0$, accurately enough to preserve this gap. This requires four properties, each an inequality between energy scales:
\begin{enumerate}
\item[(P1)] \emph{Legality}: every low-energy configuration places two ions on a diagonal of every selector.
\item[(P2)] \emph{Consistency}: every low-energy configuration has all selectors of a vertex tree on the same diagonal.
\item[(P3)] \emph{Simulation}: the energy of a consistent configuration is a constant plus $\lambda H_B$, up to the remainder in (P4).
\item[(P4)] \emph{Small remainders}: everything else, including the long-range tails, is below $\lambda/20$.
\end{enumerate}
A short logical argument (Theorem~\ref{thm:transfer}) turns these into the required yes/no equivalence. Legality follows from the size of the selectors: their half-size $b$ is polynomially small, so the internal energies $\sim b^{-1}$ dominate all interactions between selectors.

\subsection{Selectors as quadrupoles: the angular law at $s=1$}

Once every selector is legal, the configuration is described by one spin $\sigma_i=\pm1$ per selector. For two selectors with center difference $R=r(\cos\theta,\sin\theta)$, the Coulomb energy $F_R(\sigma,\sigma')$ between their four ions is a function of two binary variables, and can therefore be written exactly as~\cite{Emm26}
\begin{equation}
F_R(\sigma,\sigma')=A_R+h_R^{(1)}\sigma+h_R^{(2)}\sigma'+J_R\,\sigma\sigma' ,
\label{eq:fourier}
\end{equation}
with $J_R=\tfrac14(F_{++}-F_{+-}-F_{-+}+F_{--})$ and $h_R^{(1)}=\tfrac14(F_{++}+F_{+-}-F_{-+}-F_{--})$. Summing over pairs, the energy of any legal configuration is exactly an Ising energy: a constant, plus fields $h_i\sigma_i$, plus couplings $J_{ij}\sigma_i\sigma_j$. No multipole truncation enters this statement; the expansion is used only to evaluate and bound $h$ and $J$.

The scaling of $h$ and $J$ follows from counting multipoles. Flipping a selector leaves its charge ($2e$) and dipole moment (zero, since both diagonals are centrosymmetric) unchanged. Its leading change is in the quadrupole moment, $\Delta Q_{xy}=4b^2$ in units of the ion charge. A field $h$ is that quadrupole change interacting with the charge of another selector, so $h\sim b^2r^{-3}$; a coupling $J$ is two quadrupole changes interacting, so $J\sim b^4r^{-5}$. At $s=2$ the same counting gives $b^2r^{-4}$ and $b^4r^{-6}$~\cite{Emm26}.

\begin{lemma}[Angular law at $s=1$]
\label{lem:angular}
For unperturbed selectors with $t=b/r\le10^{-8}$,
\begin{align}
J_R&=A(\theta)\,b^4r^{-5}+\mathcal R_J,\qquad A(\theta)=\tfrac92-\tfrac{105}{2}\cos4\theta, \label{eq:J}\\
h_R^{(1)}=h_R^{(2)}&=\frac{6\sin2\theta\,b^2}{r^{3}}+\frac{20\sin2\theta\,b^4}{r^{5}}+\mathcal R_h, \label{eq:h}
\end{align}
with $|\mathcal R_J|,|\mathcal R_h|\le6\times10^4\,b^5r^{-6}$. On the axes $h_R=0$ exactly and $J_R<-47.99\,b^4r^{-5}$; at $45^\circ$, $J_R>56.99\,b^4r^{-5}$.
\end{lemma}

\begin{figure}[t]
\includegraphics[width=\columnwidth]{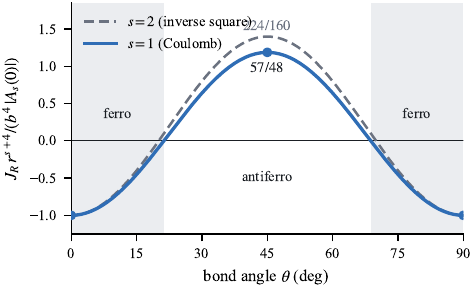}
\caption{Leading-order coupling $J_R$ between two selectors as a function of the angle $\theta$ of the vector joining them, normalized by the magnitude of its axis value: Coulomb ($s=1$, solid) and inverse square ($s=2$, dashed; Ref.~\cite{Emm26}). Shaded: the ferromagnetic sectors for the Coulomb case, used for wires; the antiferromagnetic peak at $45^\circ$ is used for edges.}
\label{fig:angular}
\end{figure}

Appendix~\ref{app:angular} derives both coefficients directly for the Coulomb potential. The quartic term in the one-selector field receives contributions from the fourth moment of one selector and from the second moments of both. Both contributions enter the field used for compensation. The sign of $A(\theta)$ is what makes the gadgets work (Fig.~\ref{fig:angular}): along the axes $J<0$ (ferromagnetic), so neighbors in a wire copy the spin, and at $45^\circ$ $J>0$ (antiferromagnetic), so the pair in an objective box anti-aligns. This sign structure, which Ref.~\cite{Emm26} found at $s=2$, survives at $s=1$, but the antiferromagnetic peak is weaker relative to the axis value ($57/48$ instead of $224/160$) and the coupling decays one power more slowly.

\subsection{Why the Coulomb tail does not spoil the encoding}
\label{sec:farfield}

The total Coulomb energy of a two-dimensional array diverges with its size: a ring of radius $r$ contains of order $r$ selectors, each contributing $\sim r^{-1}$, so every ring contributes a comparable amount. Sums of $r^{-p}$ over the plane converge only for $p>2$. The encoding nevertheless works, because only energy \emph{differences} between configurations decide the ground state, and the divergent part never enters them. The terms of Eq.~\eqref{eq:fourier} fall into four classes:
\begin{itemize}
\item \emph{Charge} ($r^{-1}$): every selector holds two ions in either state, so the monopole part of each pair energy is the same for all legal configurations. It lies entirely in the constants $A_R$, which sum to a configuration-independent offset $E_0$.
\item \emph{Dipole} ($r^{-2}$): zero in both states, so these terms vanish identically.
\item \emph{Fields} ($b^2r^{-3}$): these bias selectors toward one diagonal. Following Ref.~\cite{Emm26}, they are not bounded but \emph{cancelled}: the total field on each selector is a finite sum that the reduction computes, and stretching one diagonal of the selector by a tiny amount $\eta_i$ shifts its internal energy in one state only, by an amount chosen to cancel that sum. The same stretch at the root of each tree supplies the field term of Eq.~\eqref{eq:HB}.
\item \emph{Couplings} ($b^4r^{-5}$): these carry the logic. Couplings between unrelated gadgets are suppressed by layout: every such pair is at least a distance $D$ apart, so their total absolute coupling is at most $57.01\,b^4 M^2/(2D^5)$. With the layout parameters in Eq.~\eqref{eq:params}, $M^2/D^5=O(L^{-4})$ (Lemma~\ref{lem:remote}). Only the couplings within and near a tree require real lattice sums, and these converge like $\zeta(4)$.
\end{itemize}
This is the same structure as in Ref.~\cite{Emm26}, where the monopole sum $\sum r^{-2}$ is itself logarithmically divergent in the plane and is likewise absorbed into the constant. At $s=1$ the spin-dependent terms decay one power more slowly, which is harmless for the far field but tightens two of the near-field estimates.

\subsection{What changes at $s=1$}
\label{sec:s1}

Apart from the angular law, the proof of Ref.~\cite{Emm26} carries over step by step with rescaled constants (Appendix~\ref{app:proof}); the only parameter that must be changed is the grid scale of the layout, from $10^5L^2$ to $W=10^6L^2$. Two estimates, however, become too weak and must be replaced.

\emph{Field compensation.} The stretch $\eta_i$ that cancels the fields also perturbs the couplings. At $s=2$ the stretches are $O(Mb^5)$, and a bound on the perturbation proportional to $|\eta_i|+|\eta_j|$ suffices~\cite{Emm26}. At $s=1$ the internal energy is softer and the fields are larger, so the stretches are $O(Mb^4)$, and the same bound would allow a total coupling drift of order $M^3b^4$, far larger than $\lambda$ for every choice of $b$. The fix is again multipole counting: a stretch moves the two ions of one diagonal symmetrically apart, which leaves charge and dipole unchanged and alters only the quadrupole, by order $b\eta$.

\begin{lemma}[Perturbation sensitivity]
\label{lem:sensitivity}
For selectors at distance $r\ge1$ with $t=b/r\le10^{-8}$ and $|\eta_i|,|\eta_j|\le10^{-6}b$, $|\tilde J_{ij}-J_{ij}|\le80\,b^3r^{-5}(|\eta_i|+|\eta_j|)$ and $|\tilde h_{ij}-h_{ij}|\le10\,b\,r^{-3}(|\eta_i|+|\eta_j|)$, where tildes denote perturbed values.
\end{lemma}

\begin{figure}[t]
\includegraphics[width=\columnwidth]{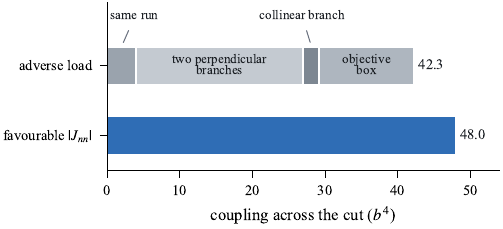}
\caption{The tightest estimate at $s=1$: the adverse coupling across the worst cut of a tree (Table~\ref{tab:budget}) against the favorable nearest-neighbor coupling, a margin of $12\%$. Applying the uniform estimate of Ref.~\cite{Emm26} with the Coulomb coefficient $|A(\theta)|\le57$ gives about $57\,b^4$ with the elementary sum estimates used there, and leaves under $2\%$ even with exact lattice sums.}
\label{fig:budget}
\end{figure}

\emph{Consistency of the trees.} A tree could carry a domain wall, a place where neighboring selectors disagree. As in Ref.~\cite{Emm26}, such a wall is removed by flipping everything on one side of it; this gains $2|J_{nn}|$ from the wall and risks at most twice the absolute coupling of all other pairs across the cut. The flip lowers the energy if the nearest-neighbor coupling outweighs the total adverse coupling across every possible cut. At $s=2$ this holds with a wide margin. At $s=1$ it is the tightest step of the proof (Fig.~\ref{fig:budget}). Using the largest angular coefficient, $|A(\theta)|\le57$, for every pair, as in the uniform estimate of Ref.~\cite{Emm26}, gives an adverse load of about $57\,b^4$ with the elementary lattice-sum estimates used there, exceeding $|J_{nn}|\approx48\,b^4$; even with exact lattice sums the uniform bound leaves a margin below $2\%$. Resolving the angular dependence of $J$ pair by pair, and certifying the resulting sums by interval arithmetic, gives the following.

\begin{lemma}[Cut budget]
\label{lem:budget}
For every cut of every tree, the absolute coupling across the cut, excluding the cut edge, is below $42.23\,b^4$, while the cut edge has $|J_{nn}|\ge47.999\,b^4$.
\end{lemma}

With Lemmas~\ref{lem:angular}--\ref{lem:budget} in place, the remaining steps (selector forcing, the objective coefficient $\lambda\approx10.5\,b^4$, remote interactions, field compensation, and a rational threshold) follow Ref.~\cite{Emm26}. Because the energies are now irrational, the threshold is chosen as a rational inside a window of width $2\lambda$ rather than computed exactly. The complete argument is in Appendix~\ref{app:proof}.

\section{Energy landscape of hard instances}
\label{sec:landscape}

The reduction also tells us what hard instances look like: they are glassy. Distinct minimizers of the source energy encode configurations that differ on entire vertex trees. Their Coulomb energies can differ slightly, but any single-ion path between them must leave the legal selector states.

\begin{corollary}[Degenerate minima behind large barriers]
\label{cor:landscape}
In the constructed instances, all minimizers of $H_B$ give configurations within $\lambda/10$ of each other in energy; every other domain-wall-free configuration lies at least $3.9\lambda$ higher, and every configuration with a domain wall at least $11\,b^4$ higher than its repaired, wall-free configuration. Distinct minimizers differ by relocating at least $2D$ ions. Any path of single-ion moves between them crosses an energy barrier of at least $0.07\,b^{-1}$, whose ratio to $\lambda$ is of order $M^{15}$. For $G$ a disjoint union of $n/4$ copies of the complete graph $K_4$, there are $10^{n/4}$ such minimizers ($4^{n/4}$ of them independent sets).
\end{corollary}

\begin{proof}
The first statements follow from Eq.~\eqref{eq:normalform} and from $2H_B\equiv n \pmod 8$ (Lemma~\ref{lem:identity}), which spaces the values of $H_B$ by 4, together with Lemma~\ref{lem:trees}. Changing the diagonal of a selector by single-ion moves passes through an illegal selector state, which by Lemma~\ref{lem:forcing} admits a single-ion move lowering the energy by more than $0.1\,b^{-1}-3M$, and so lies at least $0.1\,b^{-1}-3M$ above the global minimum. Each endpoint is within $\lambda/10$ of that minimum; hence the barrier above either endpoint exceeds $0.1\,b^{-1}-3M-\lambda/10>0.07\,b^{-1}$, by Eq.~\eqref{eq:params}. Every single-ion move from a legal configuration yields an illegal one. In each copy of $K_4$, the minimum of $q-|x|$ (Lemma~\ref{lem:identity}), namely $-1$, is attained exactly by the four single vertices and the six edges.
\end{proof}

\subsection{The cube-graph instance}
\label{sec:cube}

The cube graph in Fig.~\ref{fig:reduction} has just two maximum independent sets, its two bipartition classes. We use the corresponding ground states to illustrate the barriers in Corollary~\ref{cor:landscape}. This example illustrates the mechanism; it is not independent evidence of asymptotic hardness.

\begin{figure}[t]
\includegraphics[width=\columnwidth]{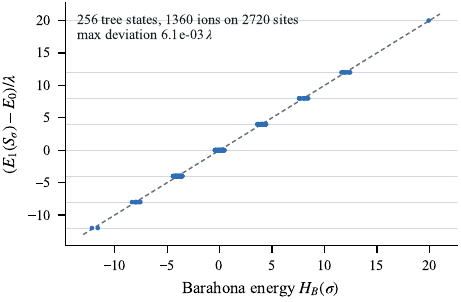}
\caption{Exact Coulomb energies of the pinned instance built from the cube graph [Fig.~\ref{fig:reduction}] for all $2^8$ domain-wall-free configurations, plotted against the source energy $H_B$ (points jittered horizontally). The energies lie within about $0.0061\lambda$ of the line $E_0+\lambda H_B$ (dashed), as Eq.~\eqref{eq:normalform} predicts.}
\label{fig:collapse}
\end{figure}

We built the instance of Fig.~\ref{fig:reduction} for the cube graph with demonstration parameters $b=0.02$, $D=8$, and grid scale $12$, much milder than those of the proof. It has 680 selectors, i.e., 1\,360 ions on 2\,720 sites. All couplings and fields were evaluated directly from Eq.~\eqref{eq:E1}, with no far-field cutoff, and the fields were compensated as in Lemma~\ref{lem:fields}. Figure~\ref{fig:collapse} shows the exact Coulomb energy of all 256 domain-wall-free configurations against $H_B$. They collapse onto the levels $E_0+\lambda H_B$ with $\lambda=10.474\,b^4$ (the objective-box sum gives $10.478\,b^4$), and the largest deviation is $0.0061\,\lambda$, well below the $\lambda/20$ the proof requires. The two ground states at $H_B=-12$ are degenerate to this accuracy.

\begin{figure}[t]
\includegraphics[width=\columnwidth]{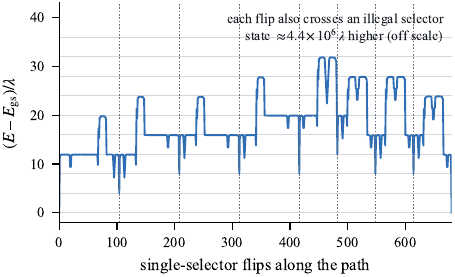}
\caption{Energy along a path of single-selector flips connecting the domain-wall-free realizations of the two source ground states. Here and in Fig.~\ref{fig:anneal}, $E_{\rm gs}$ denotes the minimum over the 256 domain-wall-free configurations. Dotted lines separate the eight trees, each flipped from a leaf inward. Horizontal lines mark multiples of $4\lambda$, the spacing of the levels of $H_B$.}
\label{fig:barrier}
\end{figure}

Along the path in Fig.~\ref{fig:barrier}, we reverse one selector at a time, starting at a leaf of each tree. Creating an isolated domain wall in a straight wire costs $2\sum_{k\ge1}k|J_k|\approx104\,b^4\approx9.9\lambda$. The path between the two ground states rises by $31.86\lambda$ when each selector is reversed in a single step.

\begin{figure}[t]
\includegraphics[width=\columnwidth]{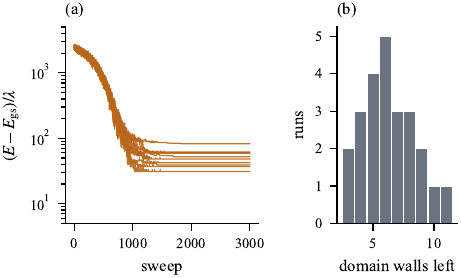}
\caption{Simulated annealing on the pinned cube-graph instance. (a) Energy above the reference minimum defined in Fig.~\ref{fig:barrier}, in units of $\lambda$, for ten runs; none reaches it. (b) Number of domain walls left in the trees at the end of each of 24 runs.}
\label{fig:anneal}
\end{figure}

In 24 simulated-annealing runs with single-selector flips, we lowered $T$ geometrically from $200\,b^4$ to $0.05\,b^4$ over 3000 sweeps (Fig.~\ref{fig:anneal}). All runs ended above this reference minimum, with 3--11 domain walls and excess energies of $24$--$82\lambda$. In a straight wire, moving a wall by one selector costs less than $0.01\,b^4$. At bends, tree roots, and objective-box lanes, a step costs $20$--$120\,b^4$, trapping the surviving walls.

\subsection{Single-ion barriers in the pinned cube}
\label{sec:barriertree-v2}
\begin{figure*}[t]
\centering
\includegraphics[width=.97\textwidth]{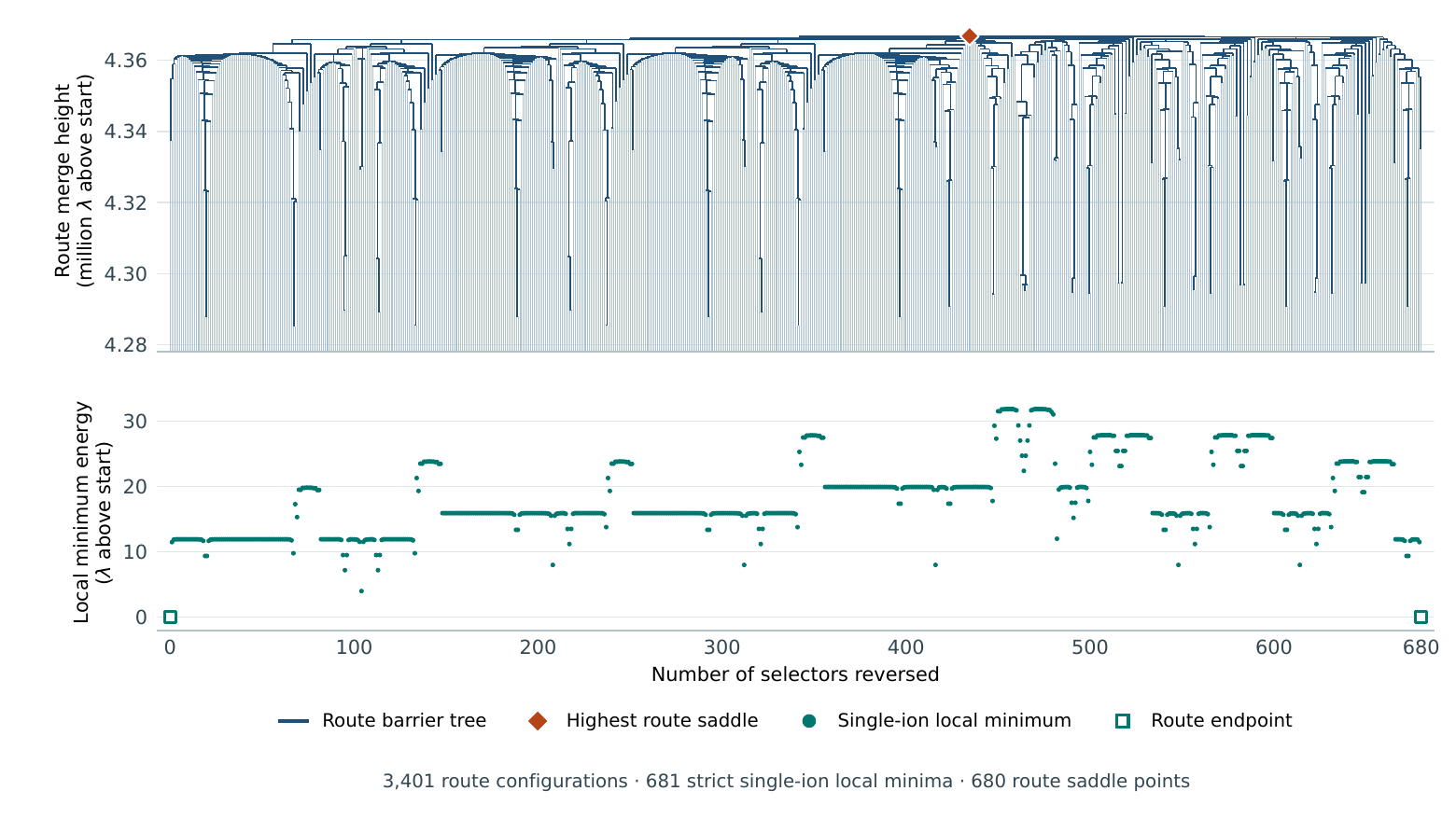}
\caption{Single-ion barrier tree for the pinned-cube route ($3\,401$ configurations). The upper panel shows the 680 route saddles in units of $10^6\lambda$. The diamond marks the highest. The lower panel shows the 681 strict local minima, each checked against every single-ion move, in units of $\lambda$. Both panels measure energy above the start. Open squares mark the endpoints. In the upper panel, two branches merge at the lowest saddle height at which the route connects them.}
\label{fig:tree-v2}
\end{figure*}

We now resolve each selector reversal in Fig.~\ref{fig:barrier} into two single-ion moves. Either ion may move first to either vacant corner. The lowest of these four intermediate energies sets the saddle for that reversal. In the barrier tree (Fig.~\ref{fig:tree-v2}), two configurations join at the lowest energy threshold that connects them~\cite{FlammBarrier2002}. On this route their joining height is the largest saddle between them.

The highest saddle, at reversal 435, gives an upper bound on the barrier between the endpoints in the full configuration graph. Every path must also visit a neighbor of each endpoint. We examined all $1\,360^2$ single-ion moves at each end and used the larger of the two lowest neighbor energies as a lower bound. Relative to the starting energy, the resulting interval is $[4.28536,\,4.36686]\times10^6\lambda$ (double precision, rounded outward). In units of $e^2/4\pi\eps_0a$, with $a$ the gadget spacing, this is $7.18$--$7.32$, and the upper end agrees to $0.1\%$ with the cost of moving one ion of a selector from a diagonal to a side, $(\tfrac12-\tfrac1{2\sqrt2})\,b^{-1}=7.32\,e^2/4\pi\eps_0a$. The barrier in units of $\lambda$ therefore scales as $0.014\,b^{-5}$, and its size of $10^6$ here reflects how small $\lambda$ is at $b=0.02$. The demonstration instance does not satisfy the hypothesis $b<0.01/M$ behind the bound $0.07\,b^{-1}$ of Corollary~\ref{cor:landscape}; for this instance the barrier is established by the computation above. The \href{https://github.com/emmerichmtm/Riesz_Hardness_Reproducibility/tree/fe25162cf565ed696bb271ef264bf449d42fbab5/notes/barrier-trees}{companion GitHub study} includes further overview figures, densities of states, numerical data, and scripts (collaborator access).

\section{Toward laboratory parameters}
\label{sec:lab}

The proof's instances are far from the laboratory. Their selectors have half-size $b\sim M^{-3}$ in units of the gadget spacing $a$, so the four sites of a selector are much closer together than any realizable pattern of wells. The proof needs this to dominate effects that grow with the instance, such as the drift of ions toward the edge of a large array, and to control the multipole remainders. To see how far the mechanism itself survives, we return to the cube-graph instance of Sec.~\ref{sec:landscape}, increase $b$, and test three numerical diagnostics using direct Coulomb-energy sums: (i) the $2^8$ tree states collapse onto $E_0+\lambda H_B$ to within $\lambda/20$; (ii) every tree edge satisfies the cut criterion of Lemma~\ref{lem:budget}, evaluated with the exact couplings and residual fields; (iii) no single-ion move from the lowest-energy domain-wall-free configuration lowers the energy. This last check establishes local stability only; it does not exclude lower-energy configurations with illegal selectors.

\begin{figure*}[t]
\includegraphics[width=\textwidth]{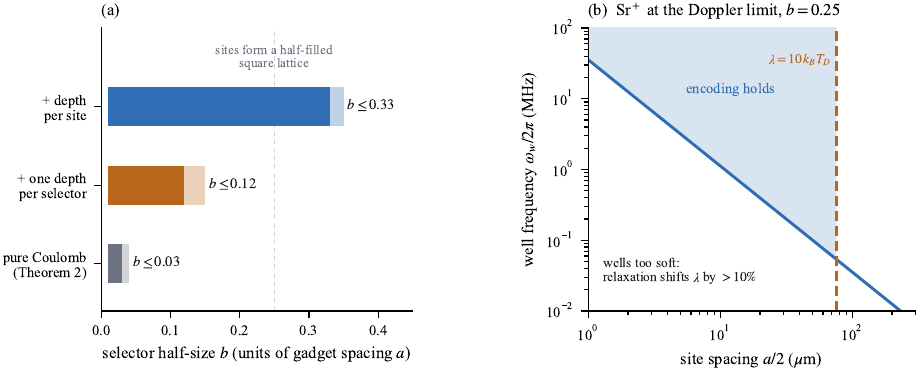}
\caption{How far toward laboratory parameters the construction survives, for the cube-graph instance (680 selectors). (a) Range of selector half-size $b$ over which the instance passes these three diagnostics [collapse within $\lambda/20$, cut criterion of Lemma~\ref{lem:budget}, and no energy-lowering single-ion move], for bare Coulomb sites as in Theorem~\ref{thm:main}, with one depth offset per selector, and with every well depth tunable. Light segments mark the first value tested at which the instance fails. At $b=0.25$ (dashed) the sites form a uniform square lattice of spacing $a/2$, half filled. (b) Laboratory window at $b=0.25$ for $^{88}$Sr$^+$ in harmonic wells at the Doppler limit $T_D\approx0.52$~mK. Above the solid line, relaxation of the ions in their wells changes $\lambda$ by less than $10\%$; left of the dashed line, $\lambda>10\,k_BT_D$.}
\label{fig:physical}
\end{figure*}

Figure~\ref{fig:physical}(a) shows the result for three levels of control over the pinning potential.
\begin{itemize}
\item \emph{Bare Coulomb sites}, as in Theorem~\ref{thm:main}: the instance works up to $b=0.03$. Beyond it, an ion lowers the energy by moving from the interior of the array to a selector near its edge.
\item \emph{One depth offset per selector}, equal to minus the mean Coulomb potential of the other selectors, with the fields cancelled by adjusting the depth of one diagonal rather than stretching it. Neither changes any coupling. The instance works up to $b=0.12$. The limit is now the local field gradient at wire ends and bends, which pushes an ion from a diagonal onto a side of its own selector.
\item \emph{Every well depth tunable}, set to cancel the mean Coulomb potential at that site: the instance works up to $b=0.33$.
\end{itemize}
Throughout, the collapse stays within $0.01\lambda$ and the cut criterion holds with a margin above $12\%$; single-ion stability fails first. At $b=0.25$ the sites form a uniform square lattice of spacing $a/2$, half filled, so with site-resolved depth control the encoding survives in a crystal-like geometry. These one-body offsets leave the Coulomb couplings untouched, but they enlarge the class of instances, and hardness with arbitrary site-dependent potentials is the easier setting discussed in Sec.~\ref{sec:intro}~\cite{WV85,Adib05}. The numerics therefore locate the physical window of the construction; they are not a hardness result at bounded spacing.

Real wells are not infinitely stiff. We model each site as a harmonic well of frequency $\omega_w$ and include, in linear response, the energy $-|F_i|^2/2m\omega_w^2$ that each ion gains by relaxing under the net force $F_i$ of the other ions, minus the mean force removed with the mean potential. Relaxation mainly renormalizes the edge coupling, which the field calibration absorbs; after recalibration the collapse stays within $0.004\lambda$. Requiring that relaxation change $\lambda$ by less than $10\%$ gives, at $b=0.25$, $\omega_w\ge16\,\omega_a$, where $\omega_a=(e^2/4\pi\eps_0ma^3)^{1/2}$; the requirement grows for smaller $b$ ($\omega_w\ge63\,\omega_a$ at $b=0.1$). Thermal energy must stay below the energy unit, $\lambda=0.047\,e^2/4\pi\eps_0a$ at $b=0.25$. For $^{88}$Sr$^+$ ions Doppler-cooled on the $422$~nm transition, $T_D=\hbar\Gamma/2k_B\approx0.52$~mK, and $\lambda>10\,k_BT_D$ for site spacings $a/2<77~\mu$m.

Figure~\ref{fig:physical}(b) shows the resulting window. At a site spacing of $5~\mu$m, $\lambda\approx150\,k_BT_D$, but the wells must reach $\omega_w\approx2\pi\times3.2$~MHz. At $20~\mu$m, $\lambda\approx40\,k_BT_D$ and $\omega_w\approx2\pi\times0.4$~MHz suffices; the rms thermal excursion is then about $90$~nm, negligible against the site spacing. The physical requirement is therefore a two-dimensional array of individually tunable, stiff wells at a pitch of a few to tens of micrometers, with Doppler cooling sufficient. This is the regime of segmented and surface-electrode trap arrays~\cite{Mielenz16}, not of optical lattices, whose wells are far too shallow (Sec.~\ref{sec:discussion}). Two caveats apply. The windows are for one instance of 680 selectors; once the mean potential is cancelled the mechanisms that close them are local, but we have not shown that the windows are independent of instance size. And the wells must also be deep enough to hold each ion against its neighbors, which harmonic wells model only near their minima.

\section{Discussion}
\label{sec:discussion}

Whether an ion-crystal ground state is hard to find in the sense of NP depends on where the structural information resides. For a fixed trap, the inputs are the unary ion count and the binary energy threshold, and the elementary P/poly bound excludes NP-hardness of this decision formulation unless the polynomial hierarchy collapses. For the fixed sphere and fixed positive-definite harmonic traps, an elementary argument also supplies polynomial-length advice for ACCS with unary ion count and precision parameter (Proposition~\ref{prop:accs}). These observations concern the input encoding and provide no uniform polynomial-time algorithm for finding configurations. The practical difficulty of free crystals is nonetheless real, approximately exponential over the range we simulated (Fig.~\ref{fig:free}), and it is the reason annealing and other global searches are used. For pinned crystals, we retain the selectors, wires, objective boxes, and field-compensation mechanism of Emmerich's inverse-square reduction~\cite{Emm26}. The new Coulomb estimates and rational-threshold construction (Sec.~\ref{sec:s1}) extend this reduction to $s=1$, establishing NP-hardness and NP-completeness of the finite-precision promise version. Our further landscape analysis gives exponentially many nearly degenerate minima for the family in Corollary~\ref{cor:landscape} and resolves barriers in the cube example using barrier trees.

Pinning potentials for ions have been realized with optical lattices superimposed on ion crystals~\cite{Linnet12,Bylinskii15}, with optical trapping of ion crystals~\cite{Schmidt18}, and with segmented-electrode arrays with many potential wells~\cite{Mielenz16}; as discussed below, only the electrode arrays currently reach the deep-pinning limit of Definition~\ref{def:crystals}(ii). Mixed-species crystals in a fixed trap are free crystals and are covered by Theorem~\ref{thm:free}. In single-molecule spectroscopy, for example, a molecular ion is held among laser-cooled atomic ions and its internal state is read out through the motion of the crystal~\cite{calvin2023single,PhysRevLett.134.153002}; where the molecular ion sits is a configurational question, but in a fixed trap it is a free-crystal problem, and for the small crystals used in these experiments it can be settled by direct enumeration. Theorems~\ref{thm:main} and~\ref{thm:npc} imply that no algorithm can be guaranteed to find the optimal loading pattern of an \emph{arbitrary} pinning landscape in polynomial time unless $\mathrm{P}=\mathrm{NP}$: on the reduction's instances, an optimal pattern together with an evaluation of its energy to precision $\lambda/2$ decides independent set.

What does the pinned limit require experimentally? The Coulomb energy of two singly charged ions a distance $a$ apart is $e^2/4\pi\eps_0a\approx k_B\times3.3$~K at $a=5~\mu$m. Definition~\ref{def:crystals}(ii) requires each well to hold its ion against its neighbors: the well depth must exceed the largest Coulomb energy an ion could gain by leaving its site. This is at least of order $e^2/4\pi\eps_0a_{\min}$ for the smallest site spacing $a_{\min}$, and for an ion that can leave the array entirely it grows to the Coulomb potential of the whole crystal at its site, of order $\sqrt{k}\,e^2/4\pi\eps_0a$ for $k$ ions at spacing $a$. The wells must also be stiff. In a well of frequency $\omega_w$ the net Coulomb force displaces each ion, which adds a configuration-dependent energy smaller than the Coulomb scale by a factor of order $(\omega_C/\omega_w)^2$, where $\omega_C=(e^2/4\pi\eps_0ma_{\min}^3)^{1/2}\approx2\pi\times0.57$~MHz for $^{88}$Sr$^+$ at $a_{\min}=5~\mu$m. Equation~\eqref{eq:E1} is the limit $\omega_w\gg\omega_C$. Electrode-defined wells in trap arrays are typically much deeper than this at spacings of tens of micrometers. Optical potentials for ions are typically much shallower than $e^2/4\pi\eps_0a$ at micrometer spacings; in current optical experiments the optical potential either corrugates a crystal held by an electrode trap or replaces the trap with a smooth confining potential, which places these experiments between the two limits or in the free one. The instances built in the proof are idealized in a further way: their selectors have sites separated by $2b\ll1$ in units of the gadget spacing, far closer than any realizable pattern of wells. Theorem~\ref{thm:main} is therefore a statement about the computational problem posed by a pinning landscape, not a claim that its hardest instances can be built in the laboratory. Section~\ref{sec:lab} shows, however, that the encoding mechanism itself survives in a half-filled square lattice of wells with tunable depths, at parameters within reach of electrode trap arrays.

Several questions remain open. Does hardness persist, as a theorem, when the site pattern has a minimum spacing comparable to the lattice constant, with or without site-dependent well depths? Does it survive for finite well depth with a harmonic envelope, where ions may leave the sites? Also open are exact membership in NP, which is tied to the sum-of-square-roots problem, and whether counting ground states is \#P-hard; the reduction does not preserve the number of solutions, because selections with internal edges can also attain $\min H_B$.

\begin{acknowledgments}
\emph{Disclosure of AI and other computational tools.}
P.~Navarro used Anthropic Claude (via Claude Code) for the
Coulomb reduction, numerical estimates, expansions, scripts,
and manuscript preparation. During the second author's
adaptation of the inverse-square reduction of
Ref.~\cite{Emm26} to the Coulomb case, Claude proposed
constants and estimates subsequently reviewed and refined
by both authors.
M.~Emmerich used OpenAI Codex for adversarial proof audits
and programming assistance, alongside Mathematica (Wolfram)
and Python for computational verification.
The research questions build on the second author's
longstanding work on the computational complexity of Riesz
energy subset selection. The results were developed
collaboratively, with all AI-generated suggestions critically
reviewed and revised by the authors, who take full
responsibility for the mathematical content.

The \href{https://github.com/Craftyleosteel/NP_Complete_Coulomb_Crystals}
{Coulomb-crystal repository} contains the code, numerical
results, Lean proofs, and a nontechnical explanation.

P.~Navarro thanks his advisor, David Patterson, for supporting this work.
\end{acknowledgments}

\appendix

\section{Proof of Theorem~\ref{thm:main} at $s=1$}
\label{app:proof}

We follow Ref.~\cite{Emm26} step by step and indicate for each step the corresponding result there. Table~\ref{tab:notation} collects the notation. Figure~\ref{fig:reduction} shows the layout used in this proof.

\begin{table}[b]
\caption{Notation for the reduction.}
\label{tab:notation}
\begin{tabular}{ll}
\toprule
Symbol & Meaning \\
\midrule
$G=(V,E)$, $n$ & planar cubic source graph, $n=|V|$ vertices \\
$\alpha(G)$, $K$ & independence number; target size \\
$x_v\in\{0,1\}$ & vertex $v$ selected ($1$) or not ($0$) \\
$\sigma_v=2x_v-1$ & the same bit as a spin, $\pm1$ \\
$H_B$, $H_0$ & source energy, Eq.~\eqref{eq:HB}; its threshold \\
$T_v$ & tree of selectors representing vertex $v$ \\
$M$, $k=2M$ & number of selectors; number of ions \\
$b$ & half-size of a selector square \\
$\eta_i$ & small enlargement of one diagonal of selector $i$ \\
$R=r(\cos\theta,\sin\theta)$ & vector between two selector centers \\
$t=b/r$ & expansion parameter, $t\le10^{-8}$ \\
$J_R$, $h_R$ & coupling and field between two selectors \\
$\lambda$ & energy unit of the encoded source problem \\
$L$, $W$, $D$ & drawing size, grid scale, feature separation \\
$E_0$ & configuration-independent energy offset \\
\bottomrule
\end{tabular}
\end{table}

\subsection{Source problem and gap transfer}
\label{app:source}

These statements are those of Ref.~\cite{Emm26}, Sec.~3. Write $x_v\in\{0,1\}$ for ``vertex $v$ selected'', $\sigma_v=2x_v-1$, $|x|=\sum_vx_v$ for the number of selected vertices, and $q$ for the number of edges with both ends selected.

\begin{lemma}[Counting identity]
\label{lem:identity}
If $G$ is cubic, then $H_B(\sigma)=n/2+4(q-|x|)$.
\end{lemma}

\begin{proof}
Substitute $\sigma_u\sigma_v=4x_ux_v-2x_u-2x_v+1$. Summing over edges gives $\sum_{uv}x_ux_v=q$, $\sum_{uv}(x_u+x_v)=3|x|$, and $|E|=3n/2$, so the edge term is $4q-6|x|+3n/2$. The field term is $2|x|-n$.
\end{proof}

\begin{lemma}[Greedy independent sets]
\label{lem:greedy}
Every set $A$ of selected vertices contains an independent set $I$ with $|I|\ge|A|-q(A)$.
\end{lemma}

\begin{proof}
We induct on $|A|$. If $A$ is independent, take $I=A$. If $A$ contains an edge $uv$, then $A'=A\setminus\{u\}$ has one vertex fewer and at least one internal edge fewer, so $|A'|-q(A')\ge|A|-q(A)$, and the independent set that the induction hypothesis gives for $A'$ also serves for $A$.
\end{proof}

\begin{corollary}[Threshold and gap]
\label{cor:gap}
Let $H_0=n/2-4K$. If $\alpha(G)\ge K$, some $\sigma$ has $H_B(\sigma)\le H_0$. If $\alpha(G)\le K-1$, every $\sigma$ has $H_B(\sigma)\ge H_0+4$.
\end{corollary}

\begin{proof}
Select an independent set of size $\ge K$, so that $q=0$ and $|x|\ge K$. Conversely, Lemma~\ref{lem:greedy} gives $|x|-q\le\alpha(G)$, and Lemma~\ref{lem:identity} gives $H_B\ge H_0+4$.
\end{proof}

The next statement isolates the logic of Ref.~\cite{Emm26}, Secs.~10--11, in a form that tolerates an irrational energy scale.

\begin{theorem}[Gap transfer]
\label{thm:transfer}
Let $\mathcal C$ be a set of configurations with energy $E$, and let $H:\{\pm1\}^V\to\Z$ be the source energy. Suppose there are reals $E_0$ and $\lambda>0$ and three maps,
\begin{itemize}
\item $\mathrm{nf}:\mathcal C\to\mathcal C$, which repairs a configuration,
\item $\mathrm{dec}:\mathcal C\to\{\pm1\}^V$, which reads off its spins,
\item $\mathrm{enc}:\{\pm1\}^V\to\mathcal C$, which builds a configuration from spins,
\end{itemize}
such that for all $c\in\mathcal C$ and $\sigma\in\{\pm1\}^V$,
\begin{align*}
E(\mathrm{nf}\,c)&\le E(c), \tag{i}\\
|E(\mathrm{nf}\,c)-E_0-\lambda H(\mathrm{dec}\,\mathrm{nf}\,c)|&<\lambda/20, \tag{ii}\\
|E(\mathrm{enc}\,\sigma)-E_0-\lambda H(\sigma)|&<\lambda/20. \tag{iii}
\end{align*}
In words: repair never raises the energy, and both repaired and encoded configurations have energy $E_0+\lambda H$ to within $\lambda/20$. Suppose further that $H$ has a gap above $H_0$: either $\min H\le H_0$ or $\min H\ge H_0+4$. Then for every threshold $\tau$ in the window
\begin{equation*}
E_0+\lambda(H_0+1)\le\tau\le E_0+\lambda(H_0+3),
\end{equation*}
the crystal problem and the source problem have the same answer:
\begin{equation*}
\exists c:\ E(c)\le\tau \iff \exists\sigma:\ H(\sigma)\le H_0 .
\end{equation*}
\end{theorem}

\begin{proof}
($\Leftarrow$) If $H(\sigma)\le H_0$, then by (iii), $E(\mathrm{enc}\,\sigma)<E_0+\lambda H_0+\lambda/20<\tau$.

($\Rightarrow$) We prove the contrapositive. If no $\sigma$ has $H(\sigma)\le H_0$, the gap gives $\min H\ge H_0+4$. Then for every $c$, by (i) and (ii),
\begin{equation*}
E(c)\ge E(\mathrm{nf}\,c)>E_0+\lambda(H_0+4)-\lambda/20>\tau .
\end{equation*}
\end{proof}

\subsection{Selector forcing}

\begin{figure}[tb]
\centering
\definecolor{ProofPlus}{HTML}{2F6DB5}%
\definecolor{ProofMinus}{HTML}{B8661A}%
\definecolor{ProofInk}{HTML}{27313B}%
\definecolor{ProofGrid}{HTML}{939CA5}%
\begin{tikzpicture}[x=1cm,y=1cm,font=\footnotesize,text=ProofInk,
  vacancy/.style={draw=ProofInk,fill=white,line width=.45pt},
  bond/.style={line width=1pt}, move/.style={->,>=stealth,draw=ProofInk,line width=.65pt}]
  % Three successive placements of two ions in the same square.
  \foreach \xc in {0,1.9,3.8}{
    \begin{scope}[shift={(\xc,0)}]
      \draw[ProofGrid,line width=.45pt] (-.46,-.46) rectangle (.46,.46);
      \foreach \xx in {-.46,.46}{\foreach \yy in {-.46,.46}
        \draw[vacancy] (\xx,\yy) circle[radius=1.65pt];}
    \end{scope}
  }
  \draw[ProofPlus,bond] (-.46,-.46) -- (.46,.46);
  \fill[ProofPlus] (-.46,-.46) circle[radius=2.1pt] (.46,.46) circle[radius=2.1pt];
  \draw[ProofInk,bond] (1.44,-.46) -- (2.36,-.46);
  \fill[ProofInk] (1.44,-.46) circle[radius=2.1pt] (2.36,-.46) circle[radius=2.1pt];
  \draw[ProofMinus,bond] (3.34,.46) -- (4.26,-.46);
  \fill[ProofMinus] (3.34,.46) circle[radius=2.1pt] (4.26,-.46) circle[radius=2.1pt];
  \draw[move] (.67,0) -- (1.22,0);
  \draw[move] (2.57,0) -- (3.12,0);
  \node at (0,-.77) {$S_+$};
  \node at (1.9,-.77) {side pair};
  \node at (3.8,-.77) {$S_-$};
  \node[anchor=west,inner sep=0pt] at (-.66,.82) {(a)};

  % Same scale within the panel; the positive stretch is exaggerated.
  \draw[ProofGrid,line width=.35pt] (4.74,-.93) -- (4.74,.9);
  \begin{scope}[shift={(5.95,0)}]
    \node[anchor=west,inner sep=0pt] at (-.78,.82) {(b)};
    \draw[ProofGrid,dashed,line width=.55pt] (-.43,-.43) rectangle (.43,.43);
    \draw[ProofPlus,bond] (-.70,-.70) -- (.70,.70);
    \draw[vacancy] (-.43,.43) circle[radius=1.65pt] (.43,-.43) circle[radius=1.65pt];
    \fill[ProofPlus] (-.70,-.70) circle[radius=2.1pt] (.70,.70) circle[radius=2.1pt];
    \node at (0,-.97) {$\eta>0$};
  \end{scope}
\end{tikzpicture}
\caption{(a) Two single-ion moves reverse an unstretched selector through a side pair. Filled circles mark ions. (b) A positive $\eta$ moves the two $S_+$ sites outward from the dashed square. The stretch is exaggerated.}
\label{fig:selector-proof}
\end{figure}
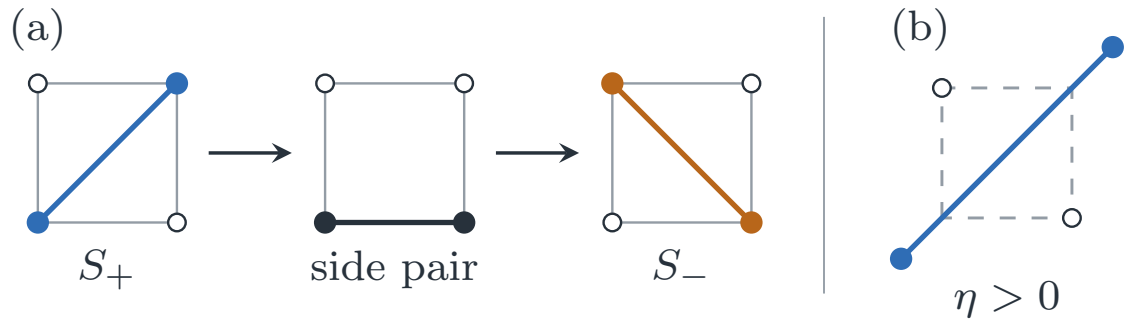

A selector with center $c\in\Z^2$ and half-size $b$ consists of the four candidate sites $c\pm(b+\eta)(1,1)$ and $c\pm b(1,-1)$. Its two legal states are the diagonals $S_+(c)=\{c\pm(b+\eta)(1,1)\}$ and $S_-(c)=\{c\pm b(1,-1)\}$. For an unstretched selector ($\eta=0$), two ions on a side cost $1/(2b)$; on a diagonal, $1/(2\sqrt2\,b)$. Figure~\ref{fig:selector-proof} shows the reversal and the stretch.

\begin{lemma}[Two-of-four forcing; cf.\ Ref.~\cite{Emm26}, Lemma~4.1]
\label{lem:forcing}
Let $M$ selectors have centers pairwise at distance $\ge1$, let $k=2M$, $|\eta_i|\le10^{-6}b$, and $b\le(40M)^{-1}$. Then every placement of $k$ ions that does not put exactly two ions on a diagonal of every selector admits a single-ion move that strictly lowers $E_1$.
\end{lemma}

\begin{proof}
Sites in distinct selectors are at distance $\ge1-2\sqrt2(b+\max|\eta|)>0.92$, so inserting an ion raises the energy between selectors by less than $2M/0.92<3M$, and removing one lowers it. If some selector holds fewer than two ions, move an ion from a selector with more than two. The minimal internal gains for the occupancy changes $(3,1)\to(2,2)$, $(3,0)\to(2,1)$, $(4,1)\to(3,2)$, and $(4,0)\to(3,1)$ are $(1-\tfrac1{2\sqrt2})b^{-1}$, $b^{-1}$, $b^{-1}$, and $(1+\tfrac1{2\sqrt2})b^{-1}$. A side pair is replaced by a diagonal with gain $(\tfrac12-\tfrac1{2\sqrt2})b^{-1}>0.146\,b^{-1}$. The enlargement $\eta$ changes each internal pair energy by less than $10^{-6}b^{-1}$, so every gain exceeds $0.1\,b^{-1}\ge4M>3M$.
\end{proof}

\subsection{Proofs of Lemmas~\ref{lem:angular} and~\ref{lem:sensitivity}}
\label{app:angular}

\begin{proof}[Proof of Lemma~\ref{lem:angular}]
Figure~\ref{fig:angular} plots the angular dependence of the coupling derived here.
Let $f(R)=1/r$, $d_\sigma=(1,\sigma)$ and $D_\sigma=d_\sigma\cdot\nabla$, with derivatives taken in the plane. For spins $\sigma,\sigma'$, the four pair energies are $f(R+b(\epsilon' d_{\sigma'}-\epsilon d_\sigma))$, with $\epsilon,\epsilon'=\pm1$. Summing over both signs removes all odd powers of $b$. The terms through degree four are
\[
\begin{aligned}
 F_R^{(\le4)}={}&4f+2b^2(D_\sigma^2+D_{\sigma'}^2)f\\
 &+b^4\!\left[\tfrac16(D_\sigma^4+D_{\sigma'}^4)
                    +D_\sigma^2D_{\sigma'}^2\right]f .
\end{aligned}
\]
In the signed average for $J_R$, only the mixed quartic term survives. Its operator is $b^4(D_+^2-D_-^2)^2/4=4b^4\partial_x^2\partial_y^2$. With $u=\sin2\theta=2xy/r^2$, this gives $(105u^2-48)b^4/r^5$, as in Eq.~\eqref{eq:J}.

For $h_R^{(1)}$ the quadratic operator is $4b^2\partial_x\partial_y$. In units of $b^4$, the pure quartic term contributes $(D_+^4-D_-^4)/12$; the mixed term contributes $(D_+^4-D_-^4)/4$. Their sum is $\tfrac83\partial_x\partial_y\Delta$, where $\Delta=\partial_x^2+\partial_y^2$. Since $\Delta f=r^{-3}$ and $\partial_x\partial_y r^{-3}=15xy/r^7$, the two quartic contributions are respectively $5u$ and $15u$ in units of $b^4/r^5$. This proves the coefficient $20u$ in Eq.~\eqref{eq:h}. Interchange of selectors gives the same field at the other end. On an axis, reflection interchanges both diagonal states and leaves all distances invariant, so each field vanishes exactly.

We bound the remainder also for use in the sensitivity estimate. Write $e_i=\eta_i/b$, $e_j=\eta_j/b$ and multiply each $+$ diagonal direction by $1+e_i$ or $1+e_j$ at its respective selector. For complex $|e_i|,|e_j|\le1/10$, each resulting displacement difference $v$ has norm at most $2.2\sqrt2$. Put $n=R/r$ and replace $t=b/r$ by $z$. A normalized pair term is $(1+w)^{-1/2}$, where $w=2z\,n\cdot v+z^2v\cdot v$ and the dot products are bilinear. On $|z|=1/10$ one has $|w|<0.72$, so each normalized Fourier coefficient has modulus at most $4/\sqrt{1-0.72}<8$. The series is even in $z$. By Cauchy's estimate its tail from degree six, in energy units, is bounded by
\[
 \frac{8}{r}\frac{(10t)^6}{1-(10t)^2}
 < \frac{9\cdot10^6}{r}t^6 .
\]
At $t\le10^{-8}$ this implies the stated remainder for the unstretched selectors.
\end{proof}

\begin{proof}[Proof of Lemma~\ref{lem:sensitivity}]
Write $a_i=1+e_i$, $a_j=1+e_j$ and $u=\sin2\theta$. In the Taylor expansion just obtained replace $D_+$ by $a_iD_+$ or $a_jD_+$ at the corresponding selector. Define the dimensionless derivatives
\[
\begin{aligned}
 P&=r^5D_+^4f=105u^2+30u-39,\\
 Q&=r^5D_+^2D_-^2f=57-105u^2 .
\end{aligned}
\]
Thus $|P|\le96$, $|Q|\le57$ and $|(P-Q)/2|\le72$ on $[-1,1]$. Let $J_4,h_4$ denote the quartic coefficients in units of $b^4/r^5$, taking the field at selector $i$. Differentiating the pure and mixed operators under the signed averages gives
\[
\begin{aligned}
 \partial_{e_i}J_4&=\tfrac12 a_i(a_j^2P-Q),\\
 \partial_{e_i}h_4&=\tfrac13a_i^3P+\tfrac12a_i(a_j^2P+Q),\\
 \partial_{e_j}h_4&=\tfrac12a_j(a_i^2P-Q).
\end{aligned}
\]
The derivative of $J_4$ in $e_j$ follows by interchanging $i,j$. For $|e_i|,|e_j|\le10^{-6}$ the coupling derivatives have magnitude below $72.01$ (write $a_j^2P-Q=(P-Q)+(a_j^2-1)P$), and both field derivatives have magnitude below $110$. The quadratic coupling vanishes for all stretches. The quadratic field in units of $b^2/r^3$ is $6u+(1+3u)(a_i^2-1)$, so either derivative has magnitude below $8.001$.

The preceding proof bounds each tail uniformly on $|e_i|,|e_j|\le1/10$. Cauchy's estimate on a circle of radius $0.09$ about either real stretch bounds its derivative in that variable by $10^8t^6/r$. Passing from $e$ to $\eta$ divides by $b$. In units of $b^3/r^5$ the coupling derivative is therefore at most $72.01+10^8t^2<73$; in units of $b/r^3$ the field derivative is at most $8.001+110t^2+10^8t^4<8.01$. Integrating the two partial derivatives from zero stretch gives the constants $73$ and $8.01$, and hence the stated bounds, also for the field at selector $j$ by symmetry.
\end{proof}

Ref.~\cite{Emm26}, Eq.~(33), uses the cruder bound $100(|\eta_i|+|\eta_j|)$ for every Fourier coefficient, which suffices at $s=2$ but not at $s=1$ (Sec.~\ref{sec:s1}).

\subsection{Layout and objective coefficient}

The gadgets are placed along a crossing-free orthogonal grid drawing of $G$ with polynomial total edge length $L_0$, computed in polynomial time~\cite{TamassiaTollis89,PT98} and scaled by $W$, exactly as in Ref.~\cite{Emm26}, Secs.~7--8. Each vertex $v$ becomes a tree $T_v$ of selectors at unit spacing along its three half-edges, and features of the drawing are kept at least $D$ apart. We set
\begin{equation}
\begin{gathered}
L=L_0+n+1,\quad W=10^6L^2,\quad D=W/100,\\
M\le10LW,\quad b=(10^8M^3)^{-1}.
\end{gathered}
\label{eq:params}
\end{equation}
Only $W$ differs from Ref.~\cite{Emm26} (there $10^5L^2$). In each objective box the two trees end in straight prefixes $p_i=(-i,0)$ and $q_j=(1+j,1)$, $0\le i,j\le D$, so that the closest pair $(p_0,q_0)$ sits at $45^\circ$ [Fig.~\ref{fig:reduction}(d)]; after its prefix, the $q$ tree returns to its corridor through a unit step. We use two properties guaranteed by the routing of Ref.~\cite{Emm26}: the only local geometries are straight runs, right-angle bends, three-branch roots, objective-box prefixes, and these unit steps; and every other pair of selectors is at distance at least $D$. The total coupling between the two prefixes defines the energy unit, $\lambda=\sum_{i,j=0}^DJ_{q_j-p_i}$.

\begin{lemma}[Objective coefficient; cf.\ Ref.~\cite{Emm26}, Lemma~8.1]
\label{lem:lambda}
$7.0\,b^4<\lambda<13.1\,b^4$, and the total absolute coupling between the two prefixes is below $13.1\,b^4$.
\end{lemma}

\begin{proof}
The distinguished pair contributes $57\cdot2^{-5/2}b^4=10.076\,b^4$ up to $10^{-3}b^4$. Every other pair has vector $(k,1)$ with $k=i+j+1\ge2$, attained by at most $k$ pairs in the finite prefixes, and $|J|\le57.01\,b^4r^{-5}$. Bounding $\sum_{k\ge2}k(k^2+1)^{-5/2}$ by its first five terms plus $\int_6^\infty x^{-4}dx$ gives $0.05234$, so the remainder is below $2.984\,b^4$.
\end{proof}

\subsection{Cut budget and tree consistency}

Flipping a set $C$ of spins changes any Ising energy by exactly $-2\sum_{i\in C,j\notin C}J_{ij}\sigma_i\sigma_j-2\sum_{i\in C}h_i\sigma_i$; this identity and the resulting gain-versus-risk criterion are machine-checked (Appendix~\ref{app:lean}).

\begin{figure}[t]
\includegraphics[width=\columnwidth]{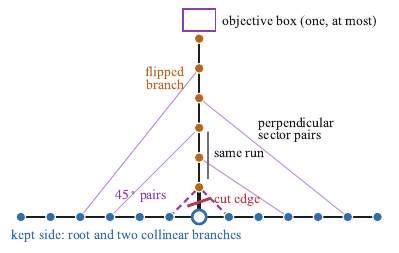}
\caption{The worst cut of a vertex tree. The cut edge (red mark) joins the root (open circle) to a branch perpendicular to both kept branches. Flipping that branch (orange) changes every coupling between it and the kept side (blue): pairs along the same line, pairs in the two perpendicular sectors (light lines; the dashed $45^\circ$ pairs nearest the root are antiferromagnetic), and at most one objective box at the end of the branch.}
\label{fig:cut}
\end{figure}

\begin{proof}[Proof of Lemma~\ref{lem:budget}]
Every pair at center separation $R=(i,j)$ satisfies $|J_R|\le w(i,j)\,b^4r^{-5}$ with $w=|A(\theta)|+6\times10^{-4}$, by Lemma~\ref{lem:angular}; for lattice vectors $|A(\theta)|$ is an exact rational, because $\cos4\theta=1-8i^2j^2/(i^2+j^2)^2$. The worst cut (Fig.~\ref{fig:cut}, with the energy comparison in Fig.~\ref{fig:budget}) joins the root of a tree to a branch perpendicular to both other branches. The pairs across it fall into the following five classes, refining the estimates of Ref.~\cite{Emm26}, Sec.~9:
\begin{itemize}
\item \emph{same straight run}: at most $k$ pairs at path distance $k\ge2$, giving at most $48.0006\,(\zeta(4)-1)$;
\item \emph{two perpendicular branches}: $2\sum_{i,j\ge1}w(i,j)(i^2+j^2)^{-5/2}$, which contains the two antiferromagnetic $45^\circ$ pairs closest to the root;
\item \emph{opposite collinear branch}: $48.0006\sum_{i,j\ge1}(i+j)^{-5}=48.0006(\zeta(4)-\zeta(5))$;
\item \emph{far end of the run}: features at path distance $\ge2D$, bounded by $2\cdot57.0006\cdot\tfrac23\sum_{i\ge2D}i^{-4}$;
\item \emph{one objective box}: bounded by Lemma~\ref{lem:lambda}.
\end{itemize}
Table~\ref{tab:budget} charges all five classes simultaneously, so it also covers the other root geometry and bends; a cut farther from the root sees only subsets of these sums. A cut at the unit step of an objective box replaces the perpendicular sectors and the collinear branch by the arm-to-arm sum $\sum_{(i,j)\ne(0,0)}w(i+j,1)\,((i+j)^2+1)^{-5/2}\le21.56$, for a total below $38.55\,b^4$. Each sum is certified by interval arithmetic with outward rounding (Appendix~\ref{app:numerics}): finite parts are summed exactly as intervals, the sector sums up to $i,j\le300$, and tails are added as intervals $[0,T]$ with analytic bounds, e.g.\ $T\le2\cdot57.0006\cdot\tfrac23\sum_{i>300}i^{-4}<3\times10^{-6}$ for the sectors. The favorable coupling satisfies $|J_{nn}|\ge(48-6\times10^{-4})\,b^4$.
\end{proof}

\begin{table}[b]
\caption{Certified upper bounds on the absolute coupling across the worst cut of a tree (Fig.~\ref{fig:cut}), in units of $b^4$. Each entry is the upper end of an interval-arithmetic enclosure that includes the multipole remainder of every pair and an analytic tail bound.}
\label{tab:budget}
\begin{tabular}{ll}
\toprule
Pairs across the cut & Certified bound \\
\midrule
same straight run & $3.9516$ \\
two perpendicular branches at the root & $23.0566$ \\
opposite collinear branch & $2.1791$ \\
far end of the run & $<10^{-9}$ \\
one objective box (Lemma~\ref{lem:lambda}) & $13.0377$ \\
total & $42.2248$ \\
\bottomrule
\end{tabular}
\end{table}

\begin{lemma}[Tree normalization; cf.\ Ref.~\cite{Emm26}, Lemma~9.1]
\label{lem:trees}
From any configuration with legal selectors, finitely many strictly energy-decreasing flips of tree branches make every tree domain-wall free, without changing the number of ions.
\end{lemma}

\begin{proof}
Root each tree at its vertex selector and flip the leaf side $C$ of a disagreeing edge $ab$. By the flip identity, the defect edge contributes $-2|J_{ab}|$, and all other terms are bounded by the cut load of Lemma~\ref{lem:budget}, the remote couplings of Lemma~\ref{lem:remote}, and the residual fields and coupling drift of Lemma~\ref{lem:fields}. In units of $b^4$,
\begin{equation}
\begin{aligned}
\frac{\Delta E}{2b^4}&\le-\frac{|J_{nn}|}{b^4}+\Lambda_{\rm cut}+\Lambda_{\rm rem}+\Lambda_{\rm field}+\Lambda_{\rm drift}\\
&\le-47.999+42.225+10^{-4}+3\times10^{-5}+10^{-5}\\
&<-5.77 .
\end{aligned}
\label{eq:budget}
\end{equation}
Every flip therefore lowers the energy by more than $11\,b^4$, and since there are finitely many configurations the process terminates.
\end{proof}

\subsection{Far field and field compensation}
\label{sec:fields}

\begin{lemma}[Remote interactions; cf.\ Ref.~\cite{Emm26}, Eq.~(46)]
\label{lem:remote}
The total absolute coupling between selector pairs not covered by Lemmas~\ref{lem:lambda} and~\ref{lem:budget} is below $10^{-4}b^4$.
\end{lemma}

\begin{proof}
All such pairs are at distance $\ge D$, there are fewer than $M^2/2$ of them, and each coupling is at most $57.01\,b^4D^{-5}$. With Eq.~\eqref{eq:params} the total is at most $2.9\times10^{13}L^2W^{-3}b^4=2.9\times10^{-5}L^{-4}b^4$.
\end{proof}

The internal field of a selector is $B(\eta)=\tfrac1{4\sqrt2}[(b+\eta)^{-1}-b^{-1}]$, with slope $\beta=-1/(4\sqrt2\,b^2)$. Let $G_i$ be the total unperturbed field on selector $i$, so $|G_i|\le6.01\,Mb^2$. The desired field $g_i$ is $\lambda$ at the root of each tree, which supplies the term $\lambda\sum_v\sigma_v$ of Eq.~\eqref{eq:HB}, and $0$ elsewhere.

\begin{lemma}[Rational field compensation; cf.\ Ref.~\cite{Emm26}, Lemma~6.1]
\label{lem:fields}
Choose rational $\eta_i$ within $\delta=10^{-4}b^6/M$ of $(g_i-G_i)/\beta$. Then $|\eta_i|\le35Mb^4$, the total residual field is below $3\times10^{-5}b^4$, the total coupling drift is below $10^{-5}b^4$, and the $\eta_i$ are computable in polynomial time.
\end{lemma}

\begin{proof}
The residual at selector $i$ consists of the curvature of $B$, at most $0.18\,b^{-3}\eta_i^2\le221M^2b^5$; the rounding, $|\beta|\delta\le1.8\times10^{-5}b^4/M$; and the field drift of Lemma~\ref{lem:sensitivity}, at most $700M^2b^5$. Here $|\eta_i|\le4\sqrt2\,b^2(\lambda+6.01Mb^2)+\delta<34.1\,Mb^4$, which also gives $|\eta_i|\le10^{-6}b$ as required by Lemmas~\ref{lem:forcing} and~\ref{lem:sensitivity}. Summing over the $M$ selectors with $M^3b=10^{-8}$, the total residual is at most $921M^3b^5+1.8\times10^{-5}b^4<2.8\times10^{-5}b^4$. The coupling drift is at most $80b^3\cdot70Mb^4\cdot M^2/2=2800M^3b^7=2.8\times10^{-5}b^6$. Computing $\eta_i$ requires $G_i$ and $\lambda$ to precision $|\beta|\delta/2$, which requires polynomially many reciprocal square roots of rationals to $O(\log M)$ bits.
\end{proof}

\subsection{Assembly}
\label{sec:proof}

Let $S_\sigma$ be the configuration that places every selector of $T_v$ on the diagonal $\sigma_v$. Expanding $E_1(S_\sigma)$ with Eq.~\eqref{eq:fourier}, the constants and the couplings within each tree (all with $\sigma_i\sigma_j=+1$) go into $E_0$, the root fields give $\lambda\sum_v\sigma_v$, and each objective box gives $\lambda\sigma_u\sigma_v$. The remaining terms are bounded by Lemmas~\ref{lem:remote} and~\ref{lem:fields}. Hence
\begin{equation}
E_1(S_\sigma)=E_0+\lambda H_B(\sigma)+\mathcal R(\sigma),
\label{eq:normalform}
\end{equation}
with $|\mathcal R|<2\times10^{-4}b^4<\lambda/20$, which establishes (P3) and (P4).

Given $(G,K)$, construct the $4M$ sites with $k=2M$ and apply Theorem~\ref{thm:transfer}: $\mathrm{nf}$ is the repair procedure of Lemmas~\ref{lem:forcing} and~\ref{lem:trees}, $\mathrm{dec}$ reads the tree spins, $\mathrm{enc}(\sigma)=S_\sigma$, and the gap is Corollary~\ref{cor:gap}. We place the threshold near $E_0+\lambda(H_0+\tfrac54)$, inside the window of Theorem~\ref{thm:transfer} and also inside the narrower window used by the Lean formalization (Appendix~\ref{app:lean}). Since $H_B(-\mathbf1)=n/2$, this point equals $E_1(S_{-\mathbf 1})+\lambda(\tfrac54-4K)-\mathcal R(-\mathbf 1)$. Output a rational $\tilde\tau$ within $\lambda/16$ of $E_1(S_{-\mathbf1})+\lambda(\tfrac54-4K)$; this takes polynomial time by evaluating each of the fewer than $2M^2$ pair terms of $E_1(S_{-\mathbf1})$ to absolute error $b^4/(64M^2)$ and $\lambda$ to absolute error $b^4/(64(4K+2))$. With $|\mathcal R(-\mathbf1)|<\lambda/20$, $\tilde\tau$ lies within $0.12\lambda$ of $E_0+\lambda(H_0+\tfrac54)$. On no-instances every configuration then has energy above $E_0+\lambda(H_0+4)-\lambda/20>\tilde\tau+1.6\lambda$, the margin used in Theorem~\ref{thm:npc}. All coordinates are rational with $O(\log M)$ bits and $M\le10^7L^3$, so the reduction runs in polynomial time. $\square$

\section{Machine-checked components}
\label{app:lean}

The following statements are proved in Lean~4 with Mathlib (source in the accompanying repository); \texttt{\#print axioms} reports only the standard axioms \texttt{propext}, \texttt{Classical.choice}, and \texttt{Quot.sound}.

\begin{itemize}
\item Lemma~\ref{lem:identity}, Lemma~\ref{lem:greedy}, and Corollary~\ref{cor:gap} (\texttt{twiceHB\_eq}, \texttt{exists\_indepSet\_large}, \texttt{barahona\_yes}, \texttt{barahona\_gap}).
\item Theorem~\ref{thm:transfer} (\texttt{Simulation.reduction\_correct}).
\item The unperturbed ($\eta=0$) internal selector inequalities of Lemma~\ref{lem:forcing} (\texttt{gain\_31\_22}, \texttt{gain\_30\_21}, \texttt{gain\_41\_32}, \texttt{gain\_40\_31}, \texttt{gain\_side\_to\_diag}, \texttt{forcing\_condition}).
\item Equation~\eqref{eq:fourier} (\texttt{two\_spin\_fourier}), the exact flip identity, and the flip-dominance criterion behind Lemma~\ref{lem:trees} (\texttt{energy\_flip}, \texttt{flip\_decreases}).
\item The assembly of Theorem~\ref{thm:main}: given a \texttt{GeometricReduction} object for $(G,K)$, the instance is a yes-instance if and only if $\alpha(G)\ge K$ (\texttt{GeometricReduction.correct}).
\end{itemize}

A \texttt{GeometricReduction} bundles a finite point set $P\subset\R^2$, $k$, a rational $\tau$, a repair map on $k$-subsets of $P$ that never raises the energy, and a simulation bound. Lean works with the integer-valued $2H_B$ (gap $8$), so its energy unit is $\lambda/2$; the required simulation bound is therefore $|\mathcal R|<\lambda/40$, and $\tau$ must lie in $[E_0+\lambda(H_0+\tfrac12),E_0+\lambda(H_0+\tfrac32)]$. The construction of Appendix~\ref{sec:proof} meets both. This is a \emph{conditional} correctness theorem: Lean does not construct the geometric object, and the geometric estimates (layout, selector forcing beyond its inequalities, Lemmas~\ref{lem:angular}--\ref{lem:budget}, \ref{lem:lambda}, \ref{lem:remote}, and~\ref{lem:fields}) are proved by hand. These are explicit inequalities in $M$, $b$, $D$, and $W$, Taylor bounds, and finite lattice sums with analytic tails, so formalizing them is routine in principle; the lattice sums, which carry the tightest margin, are already certified by interval arithmetic (Appendix~\ref{app:numerics}).

\section{Numerical methods}
\label{app:numerics}

The angular laws were obtained by symbolic expansion of the 16 ion-pair terms to sixth order in $t$ and verified against exact energies to relative accuracy $10^{-8}$ at $b=10^{-4}$.

The lattice sums of Table~\ref{tab:budget} and Lemma~\ref{lem:lambda} are certified by interval arithmetic (mpmath, 30 significant digits, outward rounding). Each pair is bounded by $(|A(\theta)|+6\times10^{-4})\,b^4r^{-5}$, with $|A(\theta)|$ evaluated as an exact rational and $r^{-5}$ enclosed as an interval. Finite parts are summed in full, the sector sums over $1\le i,j\le300$ and the one-dimensional sums to $2000$ terms. Tails are added as intervals $[0,T]$ with the analytic bounds $\sum_{k\ge n}k^{-p}\le(n-1)^{1-p}/(p-1)$ and $\sum_{j\ge1}(i^2+j^2)^{-5/2}\le\tfrac23i^{-4}$. Every enclosure has width below $10^{-5}$, and the upper ends are reported in Table~\ref{tab:budget}.

The pinned instance of Figs.~\ref{fig:collapse}--\ref{fig:anneal} uses exact pair sums of Eq.~\eqref{eq:E1} for all selector pairs and four Newton steps of field compensation. For the free crystals of Fig.~\ref{fig:free}, local minimizations used L-BFGS from Gaussian random initial positions, with minima identified by energy to relative tolerance $10^{-7}$. Simulated annealing used single-ion Metropolis moves with a geometric schedule from $T=0.05$ to $10^{-5}$ (in units of $q^2/4\pi\eps_0\ell$, where $\ell=(q^2/4\pi\eps_0m\omega^2)^{1/3}$ is the length unit of Fig.~\ref{fig:free}) over 4000 sweeps, followed by L-BFGS. Times to solution were measured as single-core wall-clock times.

For Fig.~\ref{fig:physical}, the cube-graph instance was rebuilt at each $b$ with the same layout. With bare Coulomb sites the fields were cancelled by eight Newton steps on the stretches $\eta_i$ (clipped at $|\eta_i|\le b/2$); with depth offsets, $\eta_i=0$ and the offsets were computed exactly. Collapse is the maximum deviation of the $2^8$ tree states from a linear fit in $H_B$; the cut criterion is $|J_{a v}|>\sum|J_{ij}|+\sum_{i\in C}|h_i-g_i|$ over all other pairs across the cut, for every tree edge; and legality is the minimum energy change over all single-ion moves from the ground state to an empty site. For finite stiffness, the relaxation energy was evaluated for all $2^8$ tree states and refitted with $\lambda$ and one field per tree free; the threshold stiffness is the smallest value on a logarithmic grid (5 points per decade) at which $\lambda$ changes by less than $10\%$. Scripts and saved results for the free-crystal simulations and the $b=0.02$ pinned-cube study are in the accompanying repository.

\bibliography{references,barriers_v2}

\end{document}